\documentclass[10pt,conference]{IEEEtran}

\usepackage{times,epsfig}
\usepackage{graphicx}
\usepackage{balance}
\usepackage{url}
\usepackage{leqno}
\usepackage{float}
\usepackage{algorithm}
\usepackage[noend]{algorithmic}
\usepackage{latexsym}
\usepackage{amsmath}
\usepackage{verbatim}
\usepackage{fancyvrb}
\usepackage{amssymb}
\usepackage{pifont}
\usepackage{framed}
\newtheorem{theorem}{Theorem}[section]

\newtheorem{lemma}[theorem]{Lemma}
\usepackage{verbatim}
\usepackage{graphicx}
\usepackage{balance}
\usepackage{caption}
\usepackage{subcaption}
\usepackage{url}

\usepackage{amsmath}

\usepackage[
pass,
]{geometry}

\newenvironment{myitem}{
	\begin{itemize}
		\setlength{\parskip}{0pt}
		\setlength{\itemsep}{0pt}
		\setlength{\partopsep}{0pt}
		\setlength{\parskip}{0pt}
		\setlength{\topsep}{0pt}
		\setlength{\parsep}{0pt}}{\end{itemize}
}

\newcommand{\ReachGrid}{ReachGrid}
\newcommand{\Grid}{Grid}
\newcommand{\RMBR}{RMBR}
\newcommand{\MBR}{MBR}
\newcommand{\GeoB}{GeoB}

\newcommand{\GeoReach}{{\scshape GeoReach}}
\newcommand{\SPAGraph}{{SPA-Graph}}
\newcommand{\RangeReach}{\small\sffamily RangeReach}
\newcommand{\GeoF}{GeoMT0}
\newcommand{\GeoMa}{GeoMT2}
\newcommand{\GeoMb}{GeoMT3}
\newcommand{\GeoP}{GeoP}
\newcommand{\GeoRMBR}{GeoRMBR}
\newcommand{\SpaReach}{SpaReach}
\newcommand{\MG}{\tt MAX\_REACH\_GRIDS}
\newcommand{\MR}{\tt MAX\_RMBR}
\newcommand{\Merge}{\tt MERGE\_COUNT}

\newcommand{\SWITCH}[1]{\STATE \textbf{switch} (#1)}
\newcommand{\ENDSWITCH}{\STATE \textbf{end switch}}
\newcommand{\CASE}[1]{\STATE \textbf{case} #1\textbf{:} \begin{ALC@g}}
	\newcommand{\ENDCASE}{\end{ALC@g}}

\newcommand{\DEFAULT}{\STATE \textbf{default:} \begin{ALC@g}}
	\newcommand{\ENDDEFAULT}{\end{ALC@g}}
\newcommand{\DEFAULTLINE}[1]{\STATE \textbf{default:} }

\begin{document}






%

\title{GeoReach: An Efficient Approach for Evaluating Graph Reachability Queries with Spatial Range Predicates}
%
%
%
%
%

\author{\IEEEauthorblockN{Yuhan Sun}
	\IEEEauthorblockA{CIDSE\\
		Arizona State University\\
		Tempe, AZ 85287-9309\\
		Email: Yuhan.Sun.1@asu.edu}
	\and
	\IEEEauthorblockN{Mohamed Sarwat}
	\IEEEauthorblockA{CIDSE\\
		Arizona State University\\
		Tempe, AZ 85287-9309\\
		Email:  msarwat@asu.edu}}

\maketitle
\begin{abstract}
Graphs are widely used to model data in many application domains. Thanks to the wide spread use of GPS-enabled devices, many applications assign a spatial attribute to graph vertices (e.g., geo-tagged social media). Users may issue a {\em Reachability Query with Spatial Range Predicate} ({\em abbr.} {\RangeReach}). {\RangeReach} finds whether an input vertex can reach any spatial vertex that lies within an input spatial range. An example of a {\RangeReach} query is: Given a social graph, find whether Alice can reach any of the venues located within the geographical area of Arizona State University. The paper proposes {\GeoReach} an approach that adds spatial data awareness to a graph database management system (GDBMS). {\GeoReach} allows efficient execution of {\RangeReach} queries, yet without compromising a lot on the overall system scalability (measured in terms of storage size and initialization/maintenance time). To achieve that, {\GeoReach} is equipped with a light-weight data structure, namely {\SPAGraph}, that augments the underlying graph data with spatial indexing directories. When a {\RangeReach} query is issued, the system employs a pruned-graph traversal approach.  Experiments based on real system implementation inside Neo4j proves that {\GeoReach} exhibits up to two orders of magnitude better query response time and up to four times less storage than the state-of-the-art spatial and reachability indexing approaches.

\end{abstract}

\section{Introduction}
\label{sec:introduction}

\begin{figure}[t]
  \begin{center}
  \includegraphics[width=0.44\textwidth]{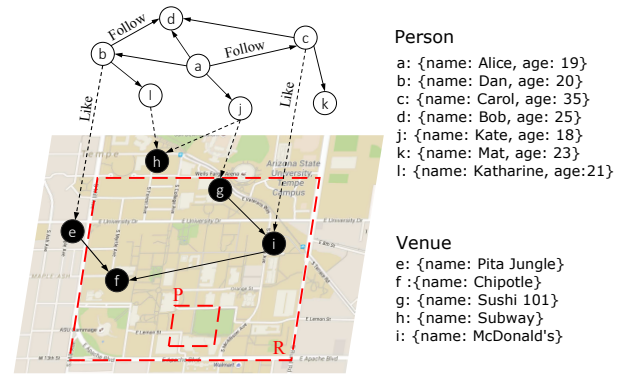}
	\caption{Location-Aware Social Graph}
  \label{fig:graph_example}
  \end{center}
\end{figure}

Graphs are widely used to model data in many application domains, including social networking, citation network analysis, studying biological function of genes, and brain simulation. A graph contains a set of vertices and a set of edges that connect these vertices. Each graph vertex or edge may possess a set of properties ({\em aka.} attributes). Thanks to the wide spread use of GPS-enabled devices, many applications assign a spatial attribute to a vertex (e.g., geo-tagged social media). Figure~\ref{fig:graph_example} depicts an example of a social graph that has two types of vertices: {\tt Person} and {\tt Venue} and two types of edges: {\tt Follow} and {\tt Like}. Vertices with type {\tt Person} have two properties (i.e., attributes): name and age. Vertices with type {\tt Venue} have two properties: name and {\em spatial} location. A spatial location attribute represents the spatial location of the entity (i.e., Venue) represented by such vertex. In Figure~\ref{fig:graph_example}, vertices $\{e,f,g,h,i\}$ are spatial vertices which represent venues.

Graph Database Management Systems (GDBMSs) emerged as a prominent NoSQL approach to store, query, and analyze graph data~\cite{GLG12,CHW+2013,SEH+13,SEH+12,SWL13}. Using a GDBMS, users can pose {\em reachability analysis} queries like: (i)~Find out whether two vertices in the graph are reachable, e.g., Are Alice (vertex $a$) and Katharine (vertex $l$) reachable in the social graph given in Figure~\ref{fig:graph_example}. (ii)~Search for graph paths that match a given regular language expression representing predicates on graph elements, e.g., Find all venues that Alice's Followees and/or her Followees' Followees also liked. Similarly, users may issue a {\em Reachability Query with Spatial Range Predicate} ({\em abbr.} {\RangeReach}). A {\RangeReach} query takes as input a graph vertex $v$ and a spatial range $R$ and returns true only if $v$ can reach any spatial vertex (that possesses a spatial attribute) which lies within the extent of $R$ (formal definition is given in Section~\ref{sec:preliminaries}). An example of a {\RangeReach} query is: Find out whether Alice can reach any of the Venues {\em located within the geographical area of Arizona State University} (depicted as a dotted red rectangle $R$ in Figure~\ref{fig:graph_example}). As given in Figure~\ref{fig:graph_example}, The answer to this query is true since Alice can reach Sushi 101 (vertex $g$) which is located within $R$. Another query example is to find out whether Katharine can reach any of the venues located within $R$. The answer to this query is false due to the fact that the only venue reachable from Katharine, Subway (vertex $h$), is not located within $R$.

There are several straightforward approaches to execute a {\RangeReach} query: (1)~{\em Traversal Approach:}~The naive approach traverses the graph, checks whether each visited vertex is a spatial vertex and returns true as the answer if the vertex's spatial attribute lies within the input query range $R$. This approach yields no storage/maintenance overhead since no pre-computed data structure is maintained. However, the Traversal approach may lead to high query response time since the algorithm may traverse the whole graph to answer the query.
(2)~{\em Transitive Closure (TC) Approach:} this approach leverages the pre-computed transitive closure~\cite{SAB+2013} of the graph to retrieve all vertices that are reachable from $v$ and returns true if at least one spatial vertex (located in the spatial range $R$) that is reachable from $v$. The TC approach achieves the lowest query response time, however it needs to pre-compute (and maintain) the graph transitive closure which is deemed notoriously infeasible especially for large-scale graphs.
(3)~{\em Spatial-Reachability Indexing (SpaReach) Approach:} uses a spatial index~\cite{BKS+90,Sam90a} to locate all spatial vertices $V_R$ that lie within the spatial range $R$ and then uses a reachability index~\cite{ZLW+2014} to find out whether $v$ can reach any vertex in $V_R$. SpaReach achieves better query response time than the Traversal approach but it still needs to necessarily probe the reachability index for spatial vertices that may never be reached from the $v$. Moreover, SpaReach has to store and maintain two index structures which may preclude the system scalability.

In this paper, we propose {\GeoReach}, a scalable and time-efficient approach that answers graph reachability queries with spatial range predicates ({\RangeReach}). {\GeoReach} is equipped with a light-weight data structure, namely {\SPAGraph}, that augments the underlying graph data with spatial indexing directories. When a {\RangeReach} query is issued, the system employs a pruned-graph traversal approach. As opposed to the SpaReach approach, {\GeoReach} leverages the Spa-Graph' s auxiliary spatial indexing information to alternate between spatial filtering and graph traversal and early prunes those graph paths that are guaranteed: (a)~not to reach any spatial vertex or (b)~to only reach spatial vertices that outside the input spatial range query. As opposed to the TC and SpaReach approaches, {\GeoReach} decides the amount of spatial indexing entries (attached to the graph) that strikes a balance between query processing efficiency on one hand and scalability (in terms of storage overhead) on the other hand. In summary, the main contributions of this paper are as follows:

\begin{itemize}

\item To the best of the authors' knowledge, the paper is the first that formally motivates and defines {\RangeReach}, a novel graph query that enriches classic graph reachability analysis queries with spatial range predicates. {\RangeReach} finds out whether an input graph vertex can reach any spatial vertex that lies within an input spatial range.


\item The paper proposes {\GeoReach} a generic approach that adds spatial data awareness to an existing GDBMS. {\GeoReach} allows efficient execution of {\RangeReach} queries issued on a GDBMS, yet without compromising a lot on the overall system scalability (measured in terms of storage size and initialization/maintenance time). 


\item The paper experimentally evaluates {\GeoReach}~\footnote{\scriptsize https://github.com/DataSystemsLab/GeoGraphDB--Neo4j} using real graph datasets based on a system implementation inside Neo4j (an open source graph database system). The experiments show that {\GeoReach} exhibits up to two orders of magnitude better query response time and occupies up to four times less storage than the state-of-the-art spatial and reachability indexing approaches.



\end{itemize}

The rest of the paper is organized as follows: Section~\ref{sec:preliminaries} lays out the preliminary background and related work. The SPA-Graph data structure, {\GeoReach} query processing, initialization and maintenance algorithms are explained in Sections~\ref{sec:georeachoverview} to~\ref{sec:initialization}. Section~\ref{sec:experiment} experimentally evaluates the performance of {\GeoReach}. Finally, Section~\ref{sec:conclusion} concludes the paper.

\section{Preliminaries and Background}
\label{sec:preliminaries}

This section highlights the necessary background and related research work. Table~\ref{table:mathnotations} summarizes the main notations in the paper.

\subsection{Preliminaries}


\begin{table} [t]
	\centering
	\begin{scriptsize}
		\renewcommand{\arraystretch}{1.2}
		\begin{tabular}{  c |  p{6cm}   }
			\hline 
			{\bf Notation} &  {\bf Description} \tabularnewline
			\hline 
			$G=\{V,E\}$ &  A graph $G$ with a set of vertices $V$ and set of edges $E$  \tabularnewline
			\hline
			$V_{v}^{out}$ & 
			The set of vertices that can be reached via a direct edge from a vertex $v$ \tabularnewline
			\hline 
			$V_{v}^{in}$ & 
			The set of vertices that can reach (via a direct edge) vertex $v$ \tabularnewline			
			\hline 
			$RF(v)$ &
			The set of vertices that are reachable from (via any number of edges) vertex $v$ \tabularnewline
			\hline
			$V_S$ & 
			The set of spatial vertices in $G$ such that $V_{S} \subseteq V$  \tabularnewline 
			\hline
			$RF_S(v)$ &
			The set of spatial vertices that are reachable from (via any number of edges) vertex $v$ \tabularnewline
			\hline
			$n$ & 
			The cardinality of V ($n=|V|$); the number of vertices in $G$  \tabularnewline 
			\hline
			$m$ & 
			The cardinality of E ($m=|E|$); the number of edges in $G$  \tabularnewline 
			\hline
			$v_1 \leadsto v_2$ & 
			$v_2$ is reachable from $v_1$ via connected path in $G$ (such that both $v_1$ and $v_2$ $\in$ $V$)  \tabularnewline 
			\hline
			{\MBR}($P$) &
			Minimum bounding rectangle of a set of spatial polygons $P$ (e.g., points, rectangles) \tabularnewline
			\hline
		\end{tabular}
	\end{scriptsize}
	\caption{Notations.}
	\label{table:mathnotations}
\end{table}

{\bf Graph Data.} {\GeoReach} deals with a directed property graph $G=(V,E)$ where (1)~$V$ is a set of vertices such that each vertex has a set of properties (attributes) and (2)~$E$ is a set of edges in which every edge can be represented as a tuple of two vertices $v_1$ and $v_2$ ($v_1,v_2\in V$). The set of spatial vertices $V_S \subseteq V$ such that each $v \in V_{S}$ has a spatial attribute (property) $v.spatial$. The spatial attribute $v.spatial$ may be a geometrical point, rectangle, or a polygon. For ease of presentation, we assume that a spatial attribute of spatial vertex is represented by a point. Figure~\ref{fig:graph_example} depicts an example of a directed property graph. Spatial Vertices $V_S$ are represented by black colored circles and are located in a two-dimensional planer space while white colored circles represent regular vertices that do not possess a spatial attribute. Arrows indicate directions of edges in the graph. 


\noindent {\bf Graph Reachability ($v_1 \leadsto v_2$).} Given two vertices $v_1$ and $v_2$ in a graph $G$, $v_1$ can reach $v_2$ ($v_1 \leadsto v_2$) or in other words $v_2$ is reachable from $v_1$ if and only if there is at least one graph path from $v_1$ to $v_2$. For example, in Figure~\ref{fig:graph_example}, vertex $a$ can reach vertex $f$ through the graph path {\small \tt a->c->i->f} so it can be represented as $a \leadsto f$. On the other hand, $c$ cannot reach $h$.

\noindent {\bf Reachability with Spatial Range Predicate ({\RangeReach}).} {\RangeReach} queries find whether a graph vertex can reach a specific spatial region (range) $R$. Given a vertex $v \in V$ in a Graph $G$ and a spatial range $R$, {\RangeReach} can be described as follows:

\begin{small}
\begin{equation}
 \emph{\small \sffamily RangeReach(v, R)} = \left\{ \,
    \begin{IEEEeqnarraybox}[][c]{l?s}
      \IEEEstrut
      {true} 	& if $\exists$~$v'$ such that\\ 
      ~	     	&		~(1)~$v' \in V_{S}$\\
      ~     	&		~(2)~$v'.spatial$ lies within $R$\\
      ~     	&		~(3)~$v \leadsto v'$\\
     {false} & Otherwise.
      \IEEEstrut
    \end{IEEEeqnarraybox}
\right.
  \label{eq:rangereach}
\end{equation}
\end{small}

As given in Equation~\ref{eq:rangereach}, if any spatial vertex $v' \in V_{S}$ that lies within the extent of the spatial range $R$ is reachable from the input vertex $v$, then {\RangeReach}($v$, $R$) returns true (i.e., $v \leadsto R$). For example, in Figure~\ref{fig:graph_example}, {\RangeReach}($a$, $R$) = \emph{true} since $a$ can reach at least one spatial vertex $f$ in $R$. However, {\RangeReach}($l$, $R$) = \emph{false} since $l$ can merely reach a spatial vertex $h$ which is not located in $R$. Vertex $d$ cannot reach $R$ since it cannot reach any vertex.

\subsection{Related Work}

This section presents previous work on reachability indexes, spatial indexes, and straightforward solutions to processing graph reachability queries with spatial range predicates ({\RangeReach}).


{\bf Reachability Index.} Existing solutions to processing graph reachability queries ($u \leadsto v$) can be divided into three categories~\cite{ZLW+2014}: (1)~Pruned Graph Traversal~\cite{CGK+2005,TSL+2007,YCZ+2010}: These approaches pre-compute some auxiliary reachability information offline. When a query is issued, the query processing algorithm traverses the graph using a classic traversal algorithm, e.g., Depth First Search (DFS) or Breadth First Search (BFS), and leverages the pre-computed reachability information to prune the search space. (2)~Transitive closure retrieval~\cite{ABJ+1989,CC+2008,J+1990,JXR+2008,SAB+2013,SSM+2011,WHY+2006}: this approach pre-computes the transitive closure of a graph offline and compresses it to reduce its storage footprint. When a query $u \leadsto v$ is posed, the transitive closure of the source vertex $u$ is fetched and decomposed. Then the query processing algorithm checks whether the terminal vertex $v$ lies in the transitive closure of $u$. and (3)~Two-Hop label matching~\cite{CPC+2010,CHW+2013,CSC+2012,CYL+2008,CHK+2003,STW+2004}: The two-hop label matching approach assigns each vertex $v$ in the graph an out-label set $L_{out}(v)$ and an in-label set $L_{in}(v)$. When a reachability query is answered, the algorithm decides that $u \leadsto v$ if and only if $L_{out}(v)\cap L_{in}(v)\not\neq\varnothing$. Since the two label sets do not contain all in and out vertices, size of the reachability index reduces.


{\bf Spatial Index.} A spatial index ~\cite{RSV02,Sam06,SC03} is used for efficient retrieval of either multi-dimensional objects (e.g., $\langle$$x$,$y$$\rangle$ coordinates of an object location) or objects with spatial extents, e.g., polygon areas represented by their minimum boundary rectangles (\MBR). Spatial index structures can be broadly classified to hierarchical (i.e., tree-based) and non-hierarchical index structures. Hierarchical tree-based spatial index structures can be classified into another two broad categories: (a)~the class of \emph{data-partitioning trees}, also known as the class of Grow-and-Post trees~\cite{Lom91}, which refers to the class of hierarchical data structures that basically extend the B-tree index structure~\cite{BM72,Com79} to support multi-dimensional and spatial objects. The main idea is to recursively partition the spatial data based on a spatial proximity clustering, which means that the spatial clusters may overlap. Examples of spatial index structures in this category include R-tree~\cite{Gut84} and R*-tree~\cite{BKS+90}. (b)~the class of \emph{space-partitioning trees} that refers to the class of hierarchical data structures that recursively decomposes the space into disjoint partitions. Examples of spatial index structures in this category include the Quad-tree~\cite{FB74} and k-d tree~\cite{Ben75}. 

{\bf Spatial Data in Graphs.} Some existing graph database systems, e.g., Neo4j, allow users to define spatial properties on graph elements. However, these systems do not provide native support for {\RangeReach} queries. Hence, users need to create both a spatial index and a reachability index to efficiently answer a {\RangeReach} queries (drawbacks of this approach are given in the following section).  On the other hand, existing research work~\cite{LMB+14} extends the RDF data with spatial data to support RDF queries with spatial predicates (including range and spatial join). However, such technique is limited to RDF and not general graph databases. It also does not provide an efficient solution to handle reachability queries. 

\subsection{Straightforward Solutions}

There are three main straightforward approaches to process a {\RangeReach} query, described as follows: 

{\bf Approach~I: Graph Traversal.} This approach executes a spatial reachability query using a classical graph traversal algorithm like DFS (Depth First Search) or BFS (Breadth First Search). When {\RangeReach}($v$, $R$) is invoked, the system traverses the graph from the starting vertex $v$. For each visited vertex, the algorithm checks whether it is a spatial vertex and returns true as the query answer if the vertex's location lies within the input query range $R$ because the requirement of spatial reachability is satisfied and hence $v \leadsto R$. Otherwise, the algorithm keeps traversing the graph. If all vertices that $v$ can reach do not lie in $R$, that means $v$ cannot reach $R$.


{\bf Approach~II: Transitive Closure (TC).} This approach pre-computes the transitive closure of the graph and stores it as an adjacency matrix in the database. Transitive closure of a graph stores the connectivity component of the graph which can be used to answer reachability query in constant time. Since the final result will be determined by spatial vertices, only spatial vertices are stored. When {\RangeReach}($v$, $R$) is invoked, the system retrieves all spatial vertices that are reachable from $v$ by means of the transitive closure. The system then returns true if at least one spatial vertex that is reachable from $v$ is also located in the spatial range $R$.

\begin{figure}[t]
	\centering
	\includegraphics[width=0.45\textwidth]{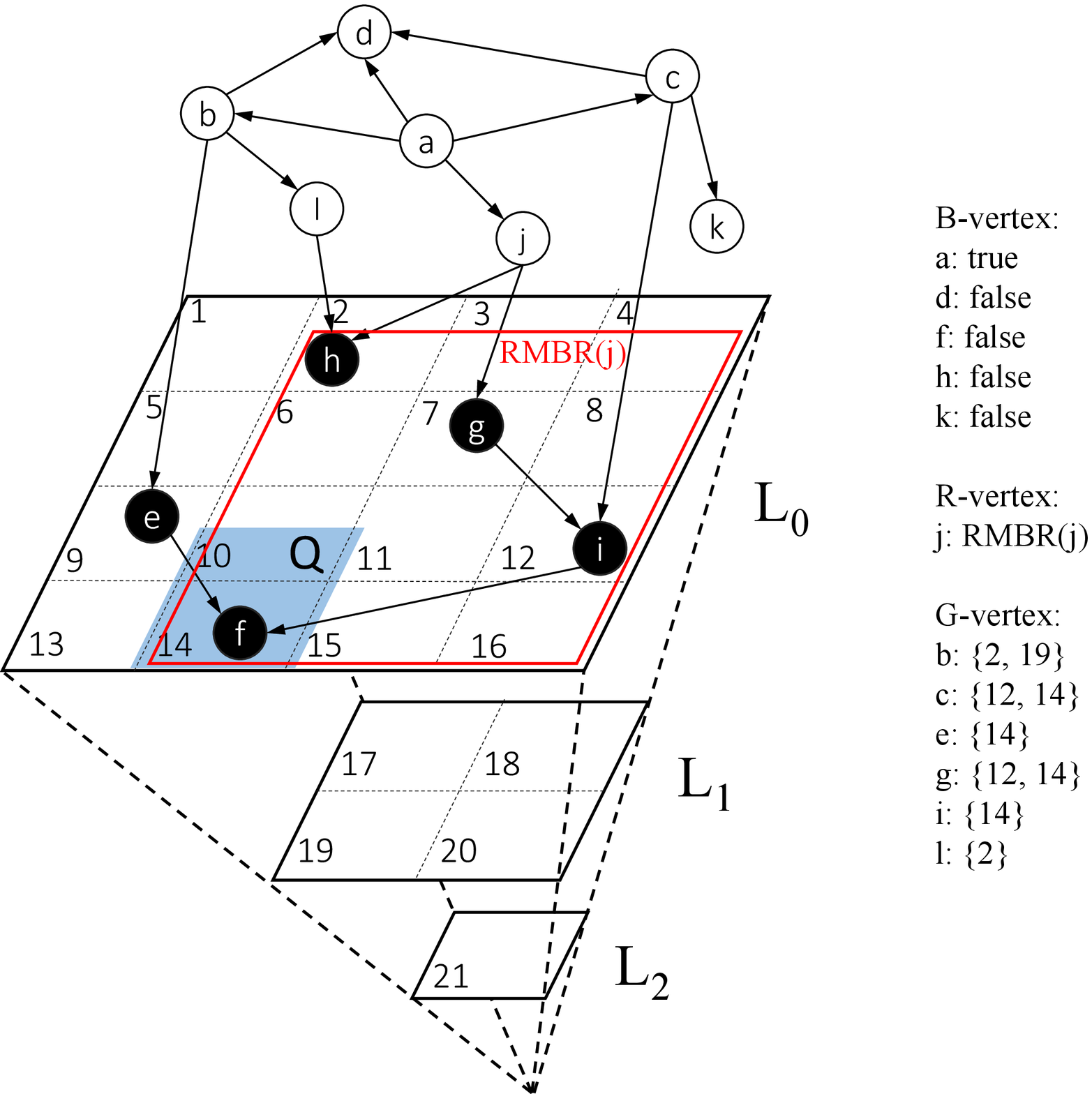}
	\caption{{\SPAGraph} Overview}
	\label{fig:graph_overview}
\end{figure}

{\bf Approach~III: SpaReach.} This approach constructs two indexes a-priori: (1)~A Spatial Index: that indexes all spatial vertices in the graph and (2)~A Reachability Index: that indexes the reachability information of all vertices in the graph. When a {\RangeReach} query is issued, the system first takes advantage of the spatial index to locate all spatial vertices $V_R$ that lie within the spatial range $R$. For each vertex $v' \in V_R$,  a reachability query against the reachability index is issued to test whether $v$ can reach $v'$. For example, to answer {\RangeReach}($a$, $R$) in Figure~\ref{fig:graph_overview}, spatial index is exploited first to retrieve all spatial vertices that are located in $R$. From the range query result, it can be known that $g$, $i$ and $f$ are located in rectangle $R$. Then graph reachability index is accessed to determine whether $a$ can reach any located-in vertex. Hence, it is obvious {\RangeReach}($a$, $R$) = \emph{true} by using this approach.


{\bf Critique.} 
The Graph Traversal approach yields no storage/maintenance overhead since no pre-computed data structure is maintained. However, the traversal approach may lead to high query response time ($O(m)$ where $m$ is the number of edges in the graph) since the algorithm may traverse the whole graph to answer the query. 
The TC approach needs to pre-compute (and maintain) the graph transitive closure which is deemed notoriously infeasible especially for large-scale graphs. The transitive closure computation is $O(kn^3)$ or $O(nm)$ and the TC storage overhead is $O(kn^2)$ where $n$ is total number of vertices and $k$ is the ratio of spatial vertices to the total number of vertices in the graph. To answer a {\RangeReach} query, the TC approach takes $O(kn)$ time since it checks whether each reachable spatial vertex in the transitive closure is located within the query rectangle. 
On the other hand, SpaReach builds a reachability index, which is a time-consuming step, in $O(n^3)$~\cite{WHY+2006} time. The storage overhead of a spatial index is $O(n)$ and that of a reachability index is $O(nm^{1/2})$. To store the two indices, the overall storage overhead is $O(nm^{1/2})$. Storage cost of this approach is far less than TC approach but still not small enough to accommodate large-scale graphs. The query time complexity of a spatial index is $O(kn)$ while that of reachability index is $m^{1/2}$. But for a graph reachability query, checking is demanded for each spatial vertex in the result set generated by the range query. Hence, cost of second step reachability query is $O(knm^{1/2})$. The total cost should be $O(knm^{1/2})$. Query performance of Spa-Reach is highly impacted by the size of the query rectangle since the query rectangle determines how many spatial vertices are located in the region. In Figure~\ref{fig:graph_example}, query rectangle $R$ overlaps with three spatial vertices. For example, to answer {\RangeReach}($l$, $R$), all three vertices $\{f,g,i\}$ will be checked against the reachability index to decide whether any of them is reachable from $l$ and in fact neither of them is reachable. In a large graph, a query rectangle will possibly contain a large number of vertices. That will definitely lead to high unreasonable high query response time.

\section{Our Approach: GeoReach}
\label{sec:georeachoverview}

In this section, we give an overview of {\GeoReach} an efficient and scalable approach for executing graph reachability queries with spatial range predicates.

\subsection{Data Structure}
\label{sec:index_overview}

In this section, we explain how {\GeoReach} augments a graph structure with spatial indexing entries to form what we call SP\lowercase{atially}-A\lowercase{ugmented} Graph ({\SPAGraph}). To be generic, \GeoReach stores the newly added spatial indexing entries the same way other properties are stored in a graph database system. The structure of a {\SPAGraph} is similar to that of the original graph except that each vertex $v \in V$ in a {\SPAGraph} $G=\{V,E\}$ stores spatial reachability information. A {\SPAGraph} has three different types of vertices, described as follows:

\begin{myitem}

\item {\bf B-Vertex:} a B-Vertex $v$ ($v \in V$) stores an extra bit (i.e., boolean), called Spatial Reachability Bit (abbr. \GeoB) that determines whether $v$ can reach any spatial vertex ($u \in V_S$) in the graph. {\GeoB} of a vertex $v$ is set to 1 (i.e., true) in case $v$ can reach at least one spatial vertex in the graph and reset to 0 (i.e., false) otherwise.

\item {\bf R-Vertex:} an R-Vertex $v$ ($v \in V$) stores an additional attribute, namely Reachability Minimum Bounding Rectangle (abbr. \RMBR($v$)). \RMBR($v$) represents the minimum bounding rectangle {\MBR}($S$) (represented by a top-left and a lower-right corner point) that encloses all spatial polygons which represent all spatial vertices $S$ that are reachable from vertex $v$ ({\RMBR}($v$) = {\MBR}($RF_S(v)$), $RF_S(v)$ = $\{u|v \leadsto u, u \in V_S\}$).

\item {\bf G-Vertex:} a G-Vertex $v$ stores a list of spatial grid cells, called the reachability grid list (abbr. {\ReachGrid}($v$)). Each grid cell $C$ in {\ReachGrid}($v$) belongs to a hierarchical grid data structure that splits the total physical space into $n$ spatial grid cells. Each spatial vertex $u \in V_S$ will be assigned a unique cell ID $(k\in [1,n])$ in case $u$ is located within the  extents of cell $k$, noted as {\Grid}$(u)=k$. Each cell $C \in$ {\ReachGrid}($v$) contains at least one spatial vertex that is reachable from $v$ ({\ReachGrid}($v$) = $\cup~${\Grid}($u$), $\{u|v\leadsto u,~{u \in {V_S}}\}$). 

\end{myitem}

\begin{lemma}
\label{lemma:RMBR}
Let $v$ ($v\in V$) be a vertex in a {\SPAGraph} $G=\{V,E\}$ and $V_{v}^{out}$ be the set of vertices that can be reached via a direct edge from a vertex $v$. The reachability minimum bounding rectangle of $v$ ({RMBR}($v$)) is equivalent to the minimum bounding rectangle that encloses all its out-edge neighbors $V_{v}^{out}$ and their reachability minimum bounding rectangles. {\small {RMBR}($v$) = {MBR}$_{v'\in V_{v}^{out}}$({RMBR}($v'$), $v'.spatial$)}.
\end{lemma}

\begin{IEEEproof}
Based on the reachability definition, the set of reachable vertices $RF(v)$ from a vertex $v$ is equal to the union of the set of vertices that is reached from $v$ via a direct edge ($V_{v}^{out}$) and all vertices that are reached from each vertex {\small$v^{'} \in V_{v}^{out}$}. Hence, the set ($RF_S(v)$) of reachable spatial vertices from $v$ is given in Equation~\ref{eq:reachablespatialvertices}.

\begin{small}
\begin{equation}
\label{eq:reachablespatialvertices}
RF_S(v) = \bigcup\limits_{v^{'}\in V_{v}^{out}} (v' \cup RF_S(v^{'}))
\end{equation} 
\end{small}

And since {\small {\RMBR}($v$) = {\MBR}($RF_S(v)$)}, then the the reachability minimum bounding rectangle of $v$ is as follows: 

\begin{small}
\begin{equation}
\begin{split}
{\RMBR}(v) & = {\MBR}(\bigcup\limits_{v^{'}\in V_{v}^{out}} (v' \cup RF_S(v^{'})))) \\
	& = {\MBR}_{v'\in V_{v}^{out}}(\RMBR(v'), v'.spatial)
\end{split}	
\end{equation}
\end{small}
That concludes the proof.
\end{IEEEproof}

\begin{lemma}
	\label{lemma:ReachGrid}
	The set of reachable spatial grid cells from a given vertex $v$ is equal to the union of all spatial grid cells reached from its all its out-edge neighbors and grid cells that contain the spatial neighbors
	\begin{small} 
	\begin{equation}
		ReachGrid(v) = \bigcup\limits_{v'\in V_v^{out}}(ReachGrid(v') \cup Grid(v'))
	\end{equation}	
	\end{small}	
\end{lemma}

\begin{IEEEproof}
\label{lemma:RMBR}
Similar to that of Lemma~\ref{lemma:RMBR}.
\end{IEEEproof}

{\bf Example.}  Figure~\ref{fig:graph_overview} gives an example of a {\SPAGraph}. {\GeoB} of vertex $b$ is set to $1$ (true) since $b$ can reach three spatial vertices $e$, $f$ and $h$. {\GeoB} for $d$ is $0$ since $d$ cannot reach any spatial vertex in the graph. Figure~\ref{fig:graph_overview} also gives an example of a Reachability Minimum Bounding Rectangle {\RMBR} of vertex $j$  (i.e., {\RMBR}($j$)). All reachable spatial vertices from $j$ are $g$, $i$, $h$ and $f$. Figure~\ref{fig:graph_overview} also depicts an example of {\ReachGrid}. There are three layers of grids, denoted as $L_0$, $L_1$, $L_2$ from top to bottom. The uppermost layer $L_0$ is split into $4\times4$ grid cells; each cell is assigned a unique id from $1$ to $16$. We denote grid cell with id $1$ as $G_1$ for brevity. The middle layer gird $L_1$ is split into four cells $G_{17}$ to $G_{20}$. Each cell in $L_1$ covers four times larger space than each cell in $L_0$. $G_{17}$ in $L_1$ covers exactly the same area of $G_1$, $G_2$, $G_5$, $G_6$ in $L_0$. The bottom layer $L_2$ contains only a single grid cell which covers all four grids in $L_1$ and represents the whole physical space. All spatial vertices reachable from vertex $a$ are located in $G_2$, $G_7$, $G_9$, $G_{12}$ and $G_{14}$, respectively. Hence, {\ReachGrid}($a$) can be $\{2,7,9,12,14\}$. Notice that vertex $e$ and $f$ are both located in $G_9$ and $G_{14}$ covered by $G_{19}$ in {\ReachGrid}($a$) can be replaced by $G_{19}$. Then, {\ReachGrid}($a$) = $\{2,7,12,19\}$. In fact, there exist more options to represent {\ReachGrid}($a$), such as $\{17,18,19,20\}$ or $\{21\}$ by merging into only a single grid cell in $L_2$. When we look into {\ReachGrid} of connected vertices, for instance $g$, {\ReachGrid}($g$) is $\{12, 14\}$ and {\ReachGrid}($i$) is $\{14\}$. It is easy to verify that {\ReachGrid}($g$) is {\ReachGrid}($i$)$\cup${\Grid}($i.spatial$), which accords with lemma~\ref{lemma:ReachGrid}.

{\bf {SPA-Graph} Intuition.} The main idea behind the {\SPAGraph} is to leverage the spatial reachability bit, reachability minimum bounding rectangle and reachability grid list stored in a B-Vertex, R-Vertex or a G-Vertex to prune graph paths that are guaranteed (or not) to satisfy both the spatial range predicate and the reachability condition. That way, {\GeoReach} cuts down the number of traversed graph vertices and edges and hence significantly reduce the overall latency of a {\small \sffamily RangeReach} query.

\subsection{Query Processing}
\label{sec:query}

\begin{algorithm}[t]
	\caption{Reachability Query with Spatial Range Predicate}	
	\begin{scriptsize}
		\label{alg:spa_reach_query}
		\begin{algorithmic}[1]			
			\STATE {\bf Function} \textsc{{RangeReach}}($v$, $R$)
			\STATE {{\bf if} v is a spatial vertex {\bf and} $v.spatial$ Lie In $R$ {\bf then} {\emph{return true}}}
			
			\STATE {Terminate $\leftarrow$ \emph{true}}
			
			\IF {$v$ is a B-vertex}
				\STATE {{\bf if} {\GeoB}($v$) = true {\bf then} Terminate $\leftarrow$ \emph{false}}
			\ELSIF {$v$ is a R-vertex}
				\STATE {{\bf if} $R$ full contains {\RMBR}($v$) {\bf then} \emph{return true}}
				\STATE {{\bf if} $R$ no overlap with {\RMBR}($v$) {\bf then} \emph{return false}}
				\STATE {Terminate $\leftarrow$ \emph{false}}	
			\ELSIF {$v$ is a G-vertex}
				\FOR {{\em each} grid $G_i\in$ {\ReachGrid}($v$)}
					\STATE {{\bf if} $R$ fully contains $G_i$ {\bf then} \emph{return true}}
					\STATE {$G_i$ partially overlaps with $R$ {\bf then} Terminate $\leftarrow$ false}
				\ENDFOR
			\ENDIF
			\IF {Terminate = \emph{false}}			
				\FOR {{\em each} vertex $v'\in V_{v}^{out}$}
				\STATE {{\bf if} \textsc{{RangeReach}}($v'$, $R$) = \emph{true} {\bf then} \emph{return true}}
				\ENDFOR
			\ENDIF
			\STATE {{\em return false}}
		\end{algorithmic}
	\end{scriptsize}
\label{algo:query}	
\end{algorithm}

This section explains the {\RangeReach} query processing algorithm. The main objective is to visit as less graph vertices and edges as possible to reduce the overall query latency. The query processing algorithm accelerates the SPA-Graph traversal procedure by pruning those graph paths that are guaranteed (or not) to satisfy the spatial reachability constraint. Algorithm~\ref{algo:query} gives pseudocode for query processing. The algorithm takes as input a graph vertex $v$ and query rectangle $R$. It then starts traversing the graph starting from $v$. For each visited vertex $v$, three cases might happen, explained as follows:

{\bf Case~I  (B-vertex):} In case {\GeoB} is false, a B-vertex cannot reach any spatial vertex and hence the algorithm stops traversing all graph paths after this vertex. Otherwise, further traversal from current B-vertex is required when {\GeoB} value is true. Line 4 to 5 in algorithm~\ref{algo:query} is for processing such case.

\begin{figure}	
	\centering
	\begin{subfigure}[t]{0.23\textwidth}
		\centering
		\includegraphics[width=0.8\linewidth]{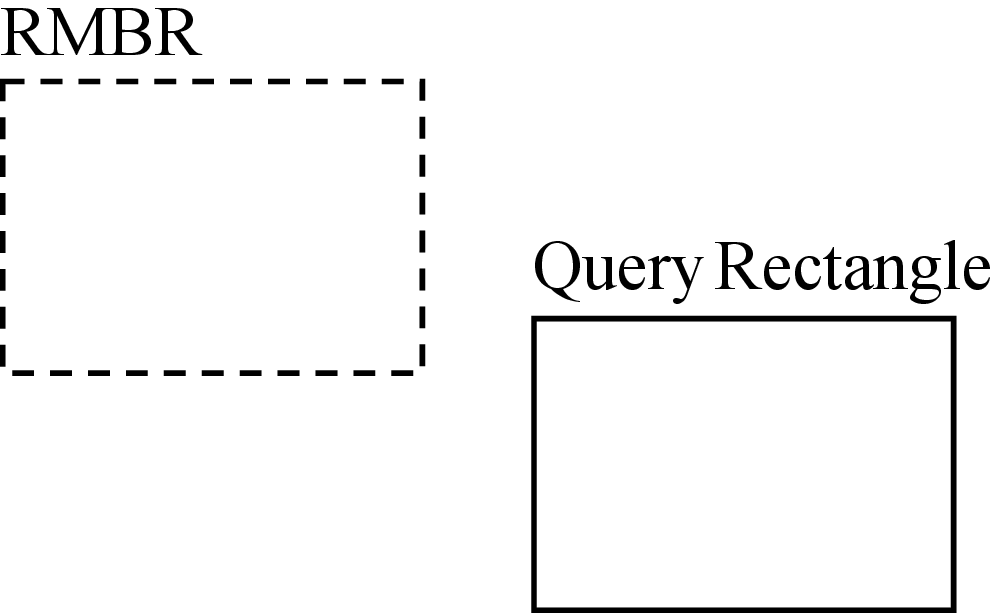}
		\caption{No Overlap}
		\label{fig:threecondition_no_overlap}
	\end{subfigure}
	\hspace{-5mm}
	\begin{subfigure}[t]{0.23\textwidth}
		\centering
		\includegraphics[width=0.7\linewidth]{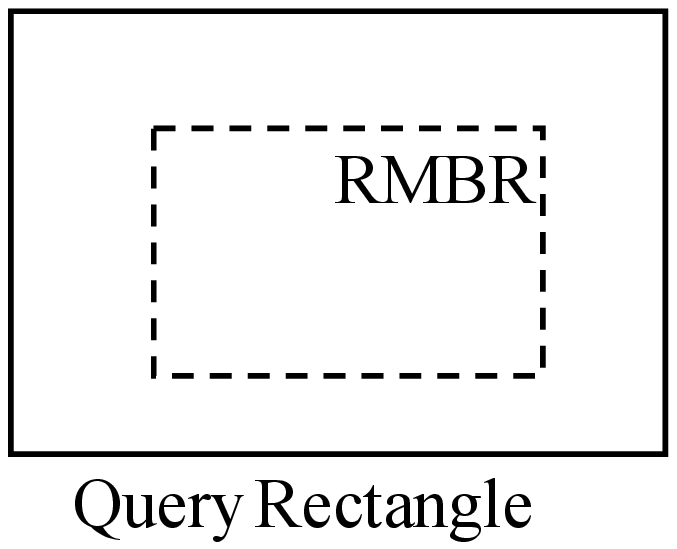}
		\caption{Lie In}
		\label{fig:threecondition_lie_in}
	\end{subfigure}
	~
	\begin{subfigure}[t]{0.46\textwidth}
		\centering
		\includegraphics[width=0.8\linewidth]{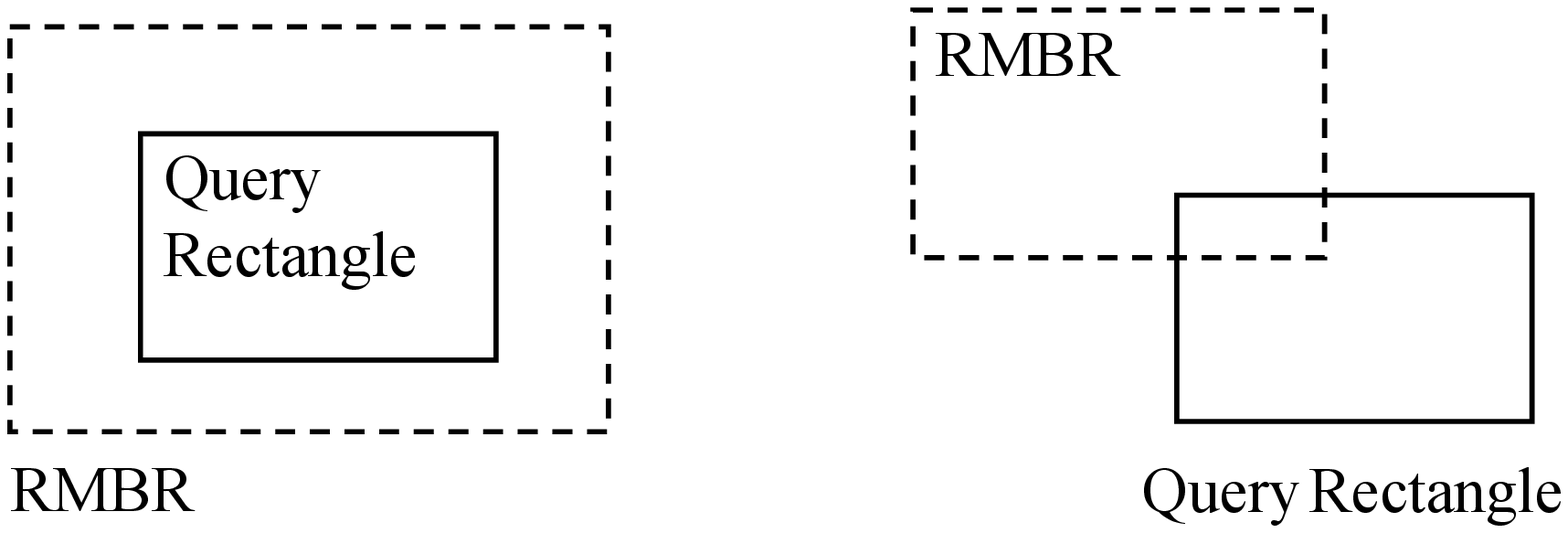}
		\caption{Partially Covered By}
		\label{fig:threecondition_partially_cover}
	\end{subfigure}
	\caption{\small Relationships between {\RMBR} and a query rectangle}
	\label{fig:threecondition}	
\end{figure}

{\bf Case~II (R-vertex):} For a visited R-vertex $u$, there are three conditions that may happen (see figure~\ref{fig:threecondition}). They are the case from line 6 to 9 in algorithm~\ref{algo:query}:
\begin{myitem}
\item {\bf Case~II.A}:  {\RMBR}($u$) lies within the query rectangle (see Figure~\ref{fig:threecondition_lie_in}). In such case, the algorithm terminates and returns {\bf true} as the answer to the query since there must exist at least a spatial vertex that is reachable from $v$.
\item {\bf Case~II.B}: The spatial query region $R$ does not overlap with {\RMBR}{($u$)} (see Figure~\ref{fig:threecondition_no_overlap}). Since all reachable spatial vertices of $u$ must lie inside {\RMBR}($u$), there is no reachable vertex can be located in the query rectangle. As a result, graph paths originating at $u$ can be pruned.
\item {\bf Case~III.C}: {\RMBR}{($u$)} is partially covered by the query rectangle (see Figure~\ref{fig:threecondition_partially_cover}). In this case, the algorithm keeps traversing the graph by fetching the set of vertices $V_{v}^{out}$ that can be reached via a direct edge from $v$.
\end{myitem}

{\bf Case~III (G-vertex):} For a G-vertex $u$, it store many reachable grids from $u$. Actually, it can be regarded as many smaller {\RMBR}s. So three cases may also happen. Algorithm~\ref{algo:query} line 13 to 18 is for such case. Three cases will happen are explained as follows: 
 \begin{myitem}
\item {\bf Case~III.A}: The query rectangle $R$ fully contains any grid cell in {\ReachGrid}{($u$)}. In such case, the algorithms terminates and returns {\em true} as the query answer.
\item {\bf Case~III.B}: The query rectangle have no overlap with all grids in {\ReachGrid}{($u$)}. This case means that $v$ cannot reach any grids overlapped with $R$. Then we never traverse from $v$ and this search branch is pruned.
\item {\bf Case~III.C}: If the query rectangle fully contains none of the reachable grid and partially overlap with any reachable grid, it corresponds to Partially Covered By case for {\RMBR}. So further traversal is performed.
 \end{myitem}

Figure~\ref{fig:graph_overview} gives an example of {\RangeReach} that finds whether vertex $a$ can reach query rectangle $Q$ (the shaded one  in figure~\ref{fig:graph_overview}). At the beginning of the traversal, the algorithm checks the category of $a$. In case, It is a B-vertex and its {\GeoB} value is true, the algorithm recursively traverses out-edge neighbors of $a$ and perform recursive checking. Therefore, the algorithm retrieves vertices $b$, $c$, $d$ and $j$. For vertex $b$, it is a G-vertex and its reachable grids are $G_2$ and $G_{19}$. $G_{19}$ cover the range of four grids in $L_0$. They are $G_9$, $G_{10}$, $G_{13}$ and $G_{14}$. The spatial range is merely partially covered by $Q$ ({\bf Case~III.C}), hence it is possible for $b$ to reach $Q$. We cannot make an assured decision in this step so $b$ is recorded for future traversal. Another neighbor is $c$. {\ReachGrid}($c$) is $\{12,14\}$ which means that $G_{12}$ and $G_{14}$ are reachable from $c$. $G_{14}$ lies in $Q$ ({\bf Case~III.A}). In such case, since $a\leadsto c$, we can conclude that $a\leadsto R$. The algorithm then halts the graph traversal at this step and returns true as the query answer.

\section{SPA-Graph Analysis}
\label{sec:spagraphanalysis}

This section analyzes each SPA-Graph vertex type rom two perspectives: (1)~Storage Overhead: the amount of storage overhead that each vertex type adds to the system (2)~Pruning Power: the probability that the query processing algorithm terminates when a vertex of such type is visited during the graph traversal.

{\bf B-vertex.} When visiting a B-Vertex, in case {\GeoB} is false, the query processing algorithm prunes all subsequent graph paths originated at such vertex. That is due to the fact that such vertex cannot reach any spatial vertex in the graph. Otherwise, the query processing algorithm continues traversing the graph. As a result, pruned power of a B-vertex lies in the condition that {\GeoB} is false. For a given graph, number of vertices that can reach any space is a certain value. So probability that a vertex can reach any spatial vertex is denoted as $P_{true}$. This is also the probability of a B-vertex whose {\GeoB} value is true. Probability of a B-vertex whose {\GeoB} value is false, denoted as $P_{false}$, will be $1-P_{true}$. To sum up, pruned power of a B-vertex is $1-P_{true}$ or $P_{false}$


\begin{figure}[t]	
	\centering
	\begin{subfigure}[t]{0.23\textwidth}
		\centering
		\includegraphics[width=1.0\linewidth]{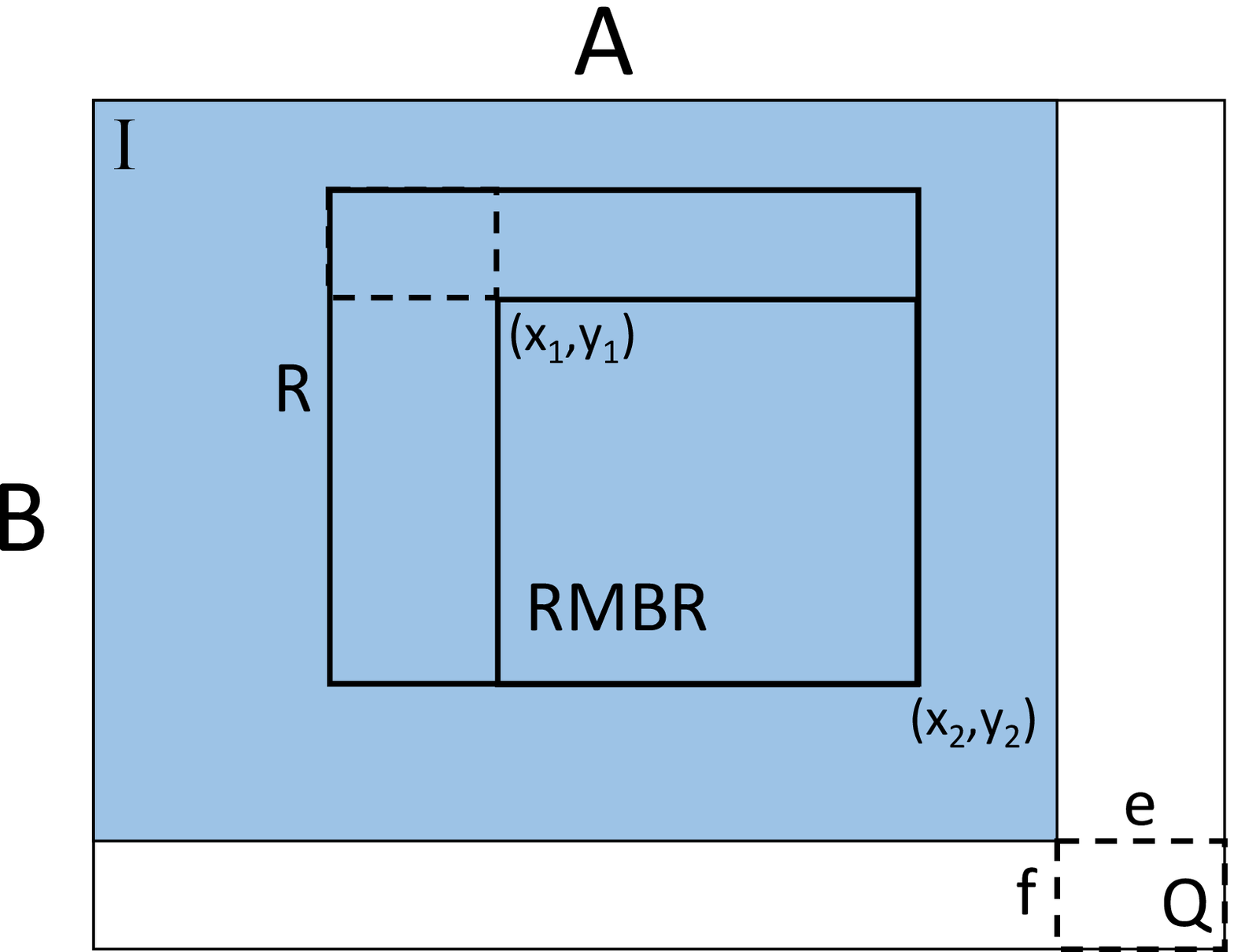}
		\caption{{}}
		\label{fig:cost_analysis_R_vertex_no_overlap}
	\end{subfigure}
	~
	\begin{subfigure}[t]{0.23\textwidth}
		\centering
		\includegraphics[width=1.0\linewidth]{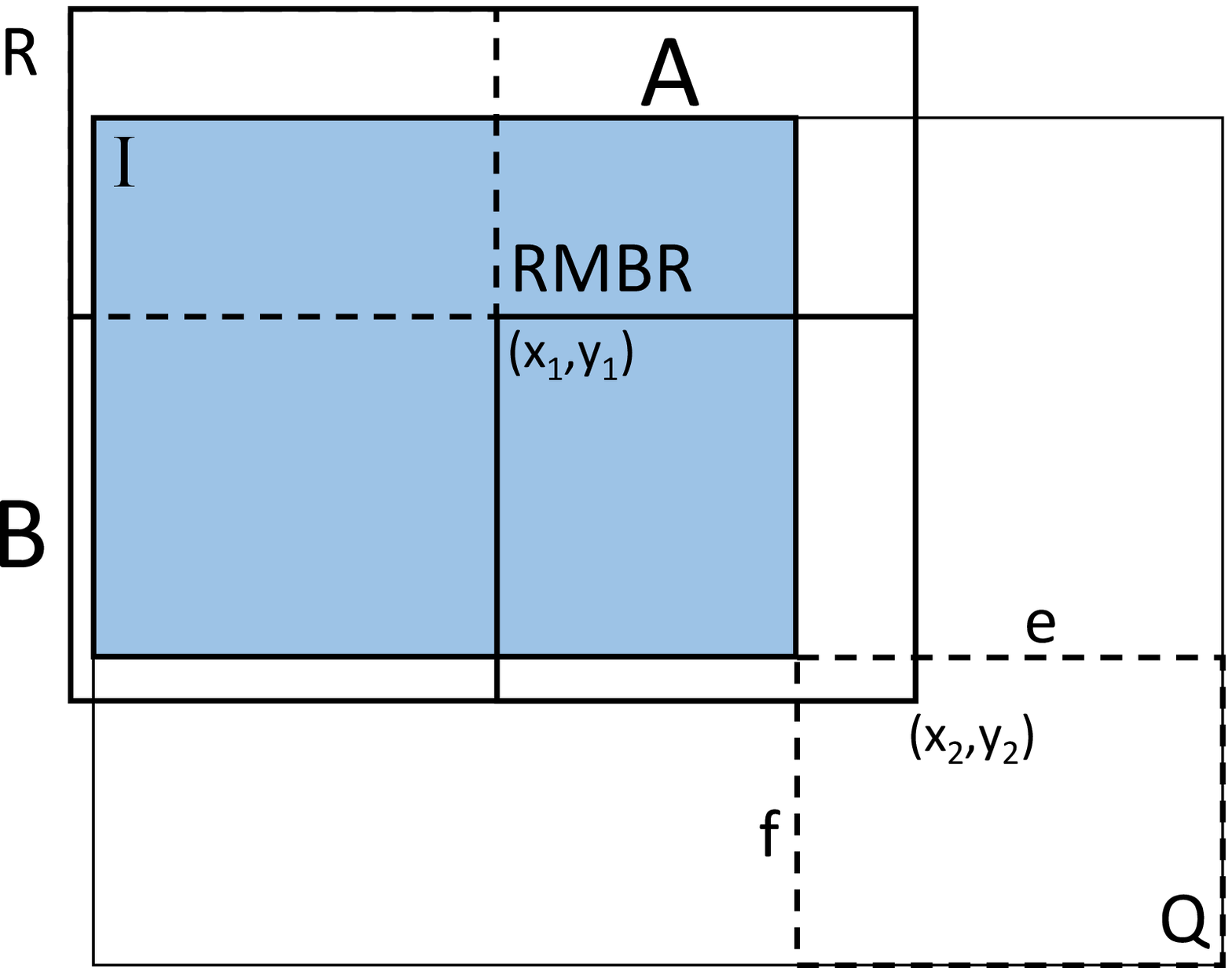}
		\caption{{}}
		\label{fig:cost_analysis_R_vertex_no_overlap_general}
	\end{subfigure}
	~
	\begin{subfigure}[t]{0.23\textwidth}
		\centering
		\includegraphics[width=1.0\linewidth]{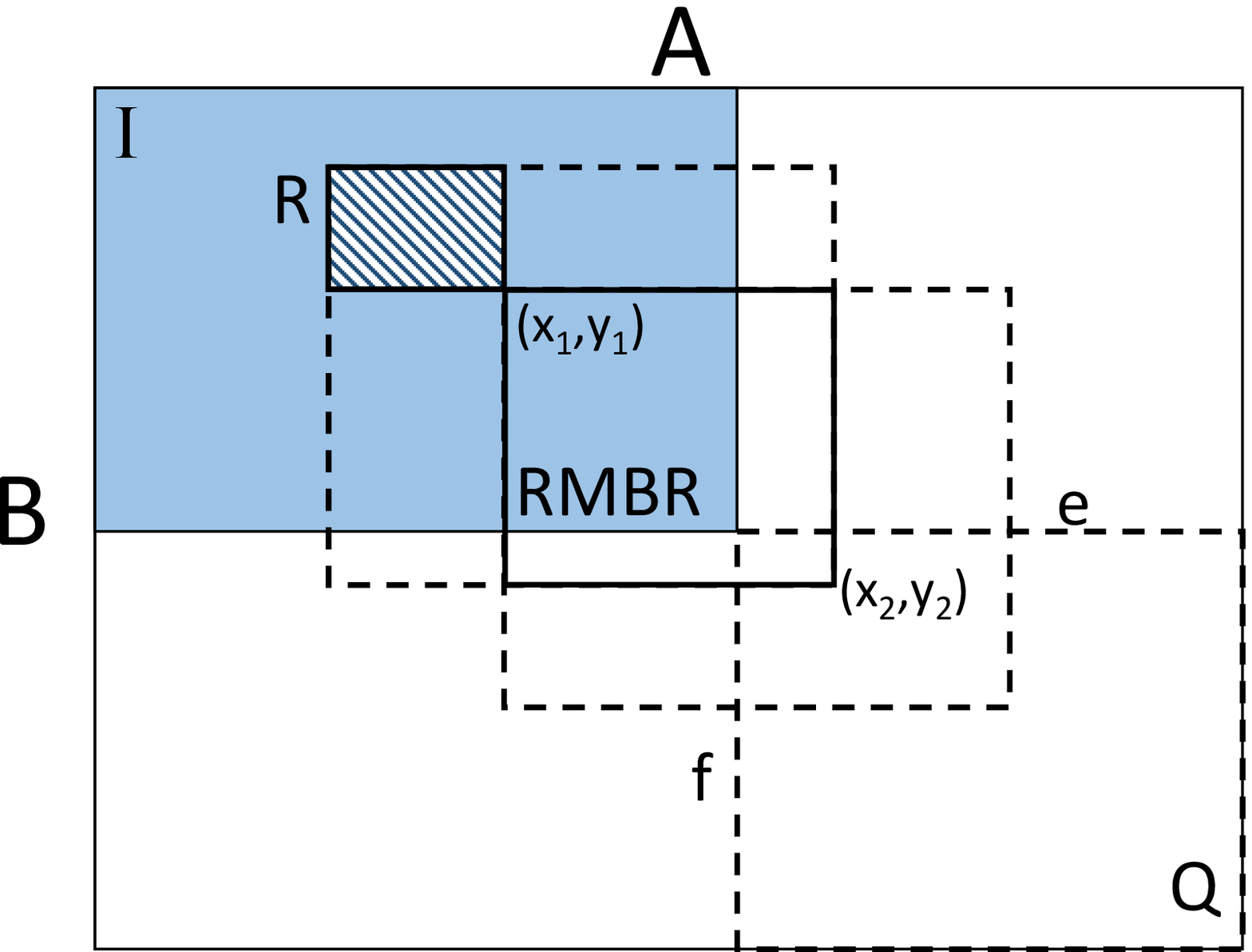}
		\caption{{}}
		\label{fig:cost_analysis_R_vertex_lie_in}
	\end{subfigure}
	\caption{R-vertex Pruning Power}\label{fig:cost_analysis_R-vertex}
	
\end{figure}

{\bf R-vertex.} When an R-vertex is visited, the condition whether the vertex can reach any space still exists. If the R-vertex cannot reach any space, we assign the R-vertex a specific value to represent it(e.g. set coordinates of {\RMBR}'s bottom-left point bigger than that of the top-right point). In this case, pruned power of a R-vertex will be the same with a B-vertex, which is $P_{false}$. Otherwise, when the R-vertex can reach some space, it will be more complex. Because information of {\RMBR} and query rectangle $R$ have some impact on the pruned power of this R-vertex. The algorithm stops traversing the graph in both the {\em No Overlap} and {\em Lie In} cases depicted in Figures~\ref{fig:threecondition_no_overlap} and~\ref{fig:threecondition_lie_in}. Figure~\ref{fig:cost_analysis_R-vertex} shows the two cases that R-vertex will stop the traversal. In Figure~\ref{fig:cost_analysis_R-vertex}, width and height of the total 2D space are denoted as $A$ and $B$. Assume that the query rectangle can be located anywhere in the space with equal probability. We use $(x_1, y_1)$ and $(x_2, y_2)$ to represent the {\RMBR}'s top-left corner and lower-right point coordinates, respectively. Then all possible areas where top-left vertex of query rectangle $Q$ should be part of the total space, denoted as $I$ (see the shadowed area in the figure. Its area is determined by size of query rectangle. Denote width and height of $Q$ are $e$ and $f$, then area of $I$, $A_I=(A-e)\times (B-f)$.

First, we estimate probability of {\em No Overlap} case. Figure~\ref{fig:cost_analysis_R_vertex_no_overlap} shows one case of {\em No Overlap}. If the query rectangle $Q$ do not overlap with {\RMBR}, top-left vertex of $Q$ must lie outside rectangle $R$ which is forms the overlap region (drawn with solid line in Figure~\ref{fig:cost_analysis_R_vertex_no_overlap_general}). Area of $R$ (denoted as $A_R$) is obviously determined by the {\RMBR} location and size of $Q$. It can be easily observed that $A_R = (x_2 - (x_1 - e))\times (y_2 - (y_1 - f))$. Another possible case is demonstrated in Figure~\ref{fig:cost_analysis_R_vertex_no_overlap_general}. In such case, if we calculate $R$ in the same way, range of $R$ will exceeds area of $I$ which contains all possible locations. As a result, $A_R = A_I$ in this case. As we can see, area of overlap region is determined by the range of $R$ and $I$ altogether. Then we can have a general representation of the overlap area $A_{Overlap} = (min(A-e, x_2) - max(0, x_1-e))\times (min(B-f, y_2) - max(0, x_2 - f)$. The {\em No Overlap} area is $A_I - A_{Overlap}$ and the probability of having a {\em No Overlap} case is calculated as follows:

\begin{equation}
P_{NoOverlap} = \frac{A_I - A_{Overlap}}{A_I} = 1 - \frac{A_{Overlap}}{A_I}.
\end{equation}

Figure~\ref{fig:cost_analysis_R_vertex_lie_in} depicts the {\em Lie In} case. When top-left vertex of $Q$ lies in region $R$, then such Lie In case will happen. To ensure that $R$ exists, it is necessary that $e>(x_2-x_1)$ and $f>(y_2-y_1)$. If it is not, then probability of such case must be 0. If this requirement is satisfied, then $A_R = (x_1 - (x_2 - e))\times (y_1 - (y_2 - f))$. Recall what is met in the above-mentioned case, $R$ may exceed the area of $I$. Similarly, more general area should be $A_R = (min(A-e, x_1) - max(0, x_1 - (x_2 - e)))\times (min(B-f, y_1) - max(0, y_1 - (y_2 - f)))$. Probability of such case should be $\frac{A_R}{A_I}$. To sum up, we have

\begin{equation}
P_{LieIn} = \begin{cases} 
\frac{A_R}{A_I}&\mbox{$e>(x_2-x_1)$ and $f>(y_2-y_1)$}\\
0&\mbox{else}
\end{cases}
\end{equation}

After we sum up all conditional probabilities based on $P_{true}$ and $P_{false}$, pruning power of an R-vertex is equal to $(P_{\small NoOverlap}+P_{\small LieIn})\times P_{true} + P_{false}$.
Evidently, the pruning power of an R-vertex is more powerful than a B-vertex. When the storage overhead of an R-vertex is considered, coordinates of {\RMBR}'s top-left and lower-right vertices should be stored. Thus its storage will be at least four bytes depending on the spatial data precision. That means the storage overhead of a G-Vertex is always higher than that of a B-Vertex.

{\bf G-vertex.}
For a high resolution grid, it is of no doubt that a G-vertex possesses a high pruning power. However, this comes at the cost of higher storage overhead because more grid cells occupies more space. When a G-vertex is compared with an R-vertex, the area of an R-vertex is much larger than a grid. In this case, an R-vertex can be seen as a a simplified G-vertex for which the grid cell size is equal to that of {\RMBR}. One extreme case of R-vertex is that the vertex can reach only one spatial vertex. In such case, {\RMBR} is location of the reachable spatial vertex. Such R-vertex can still be counted as a G-vertex whose grid size $x\rightarrow 0$. According the rule, it should be with higher storage overhead and more accuracy. Actually, storing it as a G-vertex will cost an integer while any R-vertex requires storage for four float or even double number.


\section{Initialization \& Maintenance}
\label{sec:initialization}

\begin{algorithm}[t]
	\caption{{{\GeoReach}} Initialization Algorithm}	
	\begin{scriptsize}
		\label{alg:GeoReach_construct}
		\begin{algorithmic}[1]			
			\STATE {\bf Function} \textsc{Initialize}(Graph $G=\{V,E\}$)

			\STATE {/*PHASE I: SPA-Graph Vertex Initialization */}
			\FOR {{\bf each} Vertex $v$ $\in$ $V$ according their sequence in topology}
			\STATE {{\bf InitializeVertex}($G$, $v$, {\MG}, {\MR})}
			\ENDFOR
			
			\STATE {/* PHASE II: Reachable Grid Cells Merging */}
			\label{line:phaseII:strat}
			\FOR {{\bf each} G-vertex $v$}
				\FOR {{\bf each} layer $L_i$ from $L_1$ to $L_{bottom}$}
					\FOR {{\bf each} grid cell $G_i$ in $L_i$}
						\IF {Number of reachable grids in corresponding region in $L_{i-1}$ is larger than {\Merge}}
							\STATE {Add $G_i$ in $L_i$ into {\ReachGrid}($v$)}
							\STATE {Remove reachable grid cells that are covered by $G_i$ in higher layers}
						\ENDIF
					\ENDFOR
				\ENDFOR
			\ENDFOR
			\label{line:phaseII:end}			
		\end{algorithmic}
	\end{scriptsize}
\end{algorithm}

This section describes the {\SPAGraph} initialization algorithm. The {\GeoReach} initialization algorithm (Pseudocode is given in Algorithm~\ref{alg:GeoReach_construct}) takes as input a graph Graph $G=\{V,E\}$ and runs in two main phases: {\em (1)~Phase~I: SPA-Graph Vertex Type Initialization:} this phase leverages the tradeoff between query response time and storage overhead explained in Section~\ref{sec:spagraphanalysis} to determine the type of each vertex. {\em (2)~Phase~II: Reachable Grid Cells Merging}: This step further reduces the storage overhead of each G-Vertex in the SPA-Graph by merging a set of grid cells into a single grid cell. Details of each phase are described in Section~\ref{subsec:vetextypeinit} and~\ref{subsec:cellmerging}

\subsection{SPA-Graph Vertex Type Initialization} 
\label{subsec:vetextypeinit}

\begin{algorithm}[t]
	\caption{SPA-Graph Vertex Initialization Algorithm}	
	\begin{scriptsize}
		\label{alg:GeoReach_Vertex_Init}
		\begin{algorithmic}[1]		
			\STATE {\bf Function} \textsc{InitializeVertex}(Graph $G=\{V,E\}$, Vertex $v$)
				\STATE {{\em Type} $\leftarrow$ {\bf InitializeType(v)}}		
				\SWITCH {\emph{Type}}
				
				\CASE{B-vertex}
					\STATE {Set $v$ B-vertex and {\GeoB}($v$) = \emph{true}}
				\ENDCASE
				
				\CASE{G-vertex}
					\STATE {{\ReachGrid}($v$) $\leftarrow$ $\emptyset$}
					\FOR {{\bf each} Vertex $v{'}$ $\in$ $V_{v}^{out}$}
						\STATE {\textbf{Maintain-GVertex}($v$, $v'$)}
						\IF {Number of grids in {\ReachGrid}($v$) > {\MG}}
							\STATE {Set $v$ R-vertex} and {\bf break}
						\ENDIF
					\ENDFOR
					\STATE {\emph{Type} $\leftarrow$ R-vertex}
					\IF {Number of grids in {\ReachGrid}($v$) = 0}
						\STATE {Set $v$ B-vertex, {\GeoB}($v$)} $\leftarrow$ \emph{false} and {\textbf{break}}
					\ENDIF
				\ENDCASE	
				
				\CASE{R-vertex}
					\STATE {{\RMBR}($v$) $\leftarrow$ $\emptyset$}
					\FOR {{\bf each} Vertex $v{'}$ $\in$ $V_{v}^{out}$}
						\STATE {\textbf{Maintain-RVertex}($v$, $v'$)}
						\IF {Area({\RMBR}($v$)) > {\MR}}
							\STATE {Set $v$ B-vertex, {\GeoB}($v$) $\leftarrow$ \emph{true}} and {{\bf break}}
						\ENDIF
					\ENDFOR				
				\ENDCASE
				
				\ENDSWITCH				
								
		\end{algorithmic}
	\end{scriptsize}
\end{algorithm}

To determine the type of each vertex, the initialization algorithm takes into account the following system parameters:

\begin{myitem}
\item {\bf {\MR}}: This parameter represents a threshold that limits space area of each {\RMBR}. If a vertex $v$ is an R-vertex, area of {\RMBR}($v$) cannot be larger than {\MR}. Otherwise, $v$ will be degraded to a B-vertex.
\item {\bf {\MG}}: This parameter sets up the maximum number of grid cells in each {\ReachGrid}. If a vertex $v$ is a G-vertex, number of grid cells in {\ReachGrid}($v$) cannot exceed {\MG}. Otherwise, $v$ will be degraded to an R-vertex.	
\end{myitem}

Algorithm~\ref{alg:GeoReach_Vertex_Init} gives the pseudocode of the vertex initialization algorithm. Vertices are processed based on their topological sequence in the graph. For each vertex, the algorithm first determines the initial vertex type using the {InitializeType} function (pseudocode omitted for brevity). For a vertex $v$, categories of vertex $v'$ ($\{v'|~v'\in V_{v}^{out}\}$) will be checked. If there is any B-vertex $v'$ with {\GeoB}($v'$) = \emph{true}, $v$ is directly initialized to a B-vertex with {\GeoB}($v$) = \emph{true}. Otherwise, if there is any R-vertex, the function will return an R-vertex type, which means that $v$ is initialized to R-vertex. If either of the above happens, the function returns G-vertex type. Based on the initial vertex type, the algorithm may encounter one of the following three cases:

{\bf Case~I (B-vertex):} The algorithm directly sets $v$ as a B-vertex and {\GeoB}($v$) = \emph{true} because there must exist one out-edge neighbor $v'$ of $v$ such that {\GeoB}($v'$) = \emph{true}.

{\bf Case~III (R-vertex):} For each $v'$ ($v' \in V_v^{out}$), the algorithm calls the {Maintain-RVertex} algorithm. Algorithm~\ref{alg:Ini_R} shows the pseudocode of the {Maintain-RVertex} algorithm. {Maintain-RVertex} aggregates {\RMBR} information. After each aggregation step, area of {\RMBR}($v$) will be compared with {\MR}: In case the area of {\RMBR}($v$) is larger than {\MR}, the algorithm sets $v$ to be a B-vertex with a true \GeoB value and terminates. When $v'$ is either a G-vertex or an R-vertex, the algorithm uses the new bounding rectangle returned from {\MBR}({\RMBR}($v$), {\RMBR}($v'$), $v'.spatial$) to update the current {\RMBR}($v$). The algorithm calculates the {\RMBR} of a G-vertex in case~III. In case $v'$ is a B-vertex, {\GeoB}($v'$) must be reset to false. The algorithm then updates {\RMBR}($v$) to {\MBR}({\RMBR}($v$), $v.spatial$).

\begin{algorithm}[t]
	\caption{Maintain R-vertex}	
	\begin{scriptsize}
		\label{alg:Ini_R}
		\begin{algorithmic}[1]			
			\STATE {\bf Function} \textsc{Maintain-RVertex}(From-side vertex $v$, To-side vertex $v'$)
			\SWITCH {Type of $v'$}
			
			\CASE {B-vertex}
				\IF {{\GeoB}($v'$) = true}
					\STATE {Set $v'$ B-vertex}~and~{{\GeoB}($v$) $\leftarrow$ \emph{true}}
				\ELSIF {{\RMBR}($v$) fully contains {\MBR}($v'.spatial$)}	
					\RETURN {\emph{false}}
				\ELSE
					\STATE {{\RMBR}($v$) $\leftarrow$ {\MBR}({\RMBR}($v$), $v'.spatial$)}					
				\ENDIF
			\ENDCASE
			
			\CASE {R-vertex}
				\IF {{\RMBR}($v$) fully contains {\MBR}({\RMBR}($v'$), $v'.spatial$)}
					\RETURN {\emph{false}}
				\ELSE
					\STATE {{\RMBR}($v$) $\leftarrow$ {\MBR}({\RMBR}($v$), {\RMBR}($v'$), $v'.spatial$)}
				\ENDIF
			\ENDCASE
			
			\CASE {G-vertex}
				\IF {{\RMBR}($v$) fully contains {\MBR}({\RMBR}($v'$), $v'.spatial$)}
					\RETURN {\emph{false}}
				\ELSE
					\STATE {{\RMBR}($v$) $\leftarrow$ {\MBR}({\RMBR}($v$), {\RMBR}($v'$), $v'.spatial$)}
				\ENDIF
			\ENDCASE
						
			\ENDSWITCH
			\RETURN {\emph{true}}
						
		\end{algorithmic}
	\end{scriptsize}
\end{algorithm}
		
{\bf Case~II (G-vertex):} For each vertex $v'$ ($v' \in V_v^{out}$), {Maintain-GVertex} (pseudocode omitted for the sake of space) is invoked to calculate the {\ReachGrid} of $v'$. 
In case $v'$ is a B-vertex with {\GeoB}($v'$) = \emph{false} and $v'$ is a spatial vertex, the grid cell that contains the location of $v'$ will be added into {\ReachGrid}($v$). If $v'$ is a G-vertex, all grid cells in {\ReachGrid}($v'$) and {\Grid}($v'.spatial$) will be added into {\ReachGrid}($v$). It does not matter whether $v'$ is a spatial vertex or not. If $v'$ is not a spatial vertex, {\Grid}($v'.spatial$) is $\emptyset$. After accumulating information from each neighbor $v'$, the algorithm changes the type of $v$ to R-vertex immediately in case the number of reachable grid cells in {\ReachGrid}($v$) is larger than {\MG}. Therefore, the algorithm sets the \emph{Type} to R-vertex since {\RMBR}($v$) should be calculated for possible future usage, e.g. {\RMBR} of in-edge neighbors of $v$(it will be shown in R-vertex case). 

{\bf Example.} Figure~\ref{fig:graph_overview} depicts a {\SPAGraph} with {\MR} = 0.8A and {\MG} = 4, where A is area of the whole space. Each vertex is attached with some information and affiliated to one category of {{\GeoReach}} index. Their affiliations are listed in the figure. It is obvious that those vertices which cannot reach any spatial vertices will be stored as B-vertex and have a false boolean {\GeoB} value to represent such condition. Vertices $d$, $f$, $h$, $i$, $j$ and $k$ are assigned a false value. Other vertices are G-vertex initially. {\ReachGrid}($a$) = $\{2,7,9,12,14\}$, {\ReachGrid}($b$) = $\{2,9,14\}$, {\ReachGrid}($c$) = $\{12, 14\}$, {\ReachGrid}($e$) = $\{14\}$, {\ReachGrid}($g$) = $\{12,14\}$, {\ReachGrid}($i$) = $\{14\}$, {\ReachGrid}($j$) = $\{2,7,12,14\}$ and {\ReachGrid}($l$) = $\{2\}$. Because of {\MG}, some of them will be degraded to an R-vertex. Number of reachable grids in {\ReachGrid}($a$) and {\ReachGrid}($j$) are 4 and 5, respectively. Both of them are larger than or equal to {\Merge}. They will be degraded to R-vertex first. Then area of their {\RMBR} are compared with {\MR}. Area of {\RMBR}($a$) is apparently over 80\% of the total space area. According to {\MR}, $a$ is stored as a B-vertex with a true value while $j$ is stored as an R-vertex with an {\RMBR}.

\subsection{Reachable Grid Cells Merging}
\label{subsec:cellmerging} 

After the type of each vertex is decided, the initialization algorithm performs the reachable grid cells merging phase (lines~\ref{line:phaseII:strat} to ~\ref{line:phaseII:end} in Algorithm~\ref{alg:GeoReach_construct}). In this phase, the algorithm merges adjacent grid cells to reduce the overall storage overhead of each G-Vertex. To achieve that, the algorithm assumes a system parameter, namely {\bf {\Merge}}. This parameter determines how {\GeoReach} merges spatially adjacent grid cells according to {\Merge}. In each spatial region with four grid cells, the number of reachable grid cells should not be less than {\Merge}. Otherwise, we merge the four grid cells into a single grid cell in the lower layer.


For each G-vertex $v$, all grid cells in grid cell layers $L_1$ to $L_{bottom}$ are checked. When a grid cell $G_i$ in $L_i$ is processed, four grid cells in $L_{i-1}$ that cover the same space with $G_i$ will be accessed. If number of reachable grid cells is larger than or equal to {\Merge}, $G_i$ should be added in {\ReachGrid}($v$) first. Then all grid cells covered by $G_i$ in layers from $L_0$ to $L_{i-1}$ should be removed. In order to achieve that, a recursive approach is implemented as follows. For each grid cell in $L_{i-1}$ that is reachable from $v$, the algorithm directly remove it from {\ReachGrid}($v$). The removal stops at this grid in this layer. No recursive checking is required on grid cells in higher layers for which the space is covered by the reachable grid cell. Since all those reachable grid cells have been removed already. For those grid cells that are not reachable from $v$, the algorithm cannot assure that they do not cover some reachable grids in a higher layer. Hence, the recursive removal is invoked until the algorithm reaches the highest layer or other reachable grid cells are visited.


The {\SPAGraph} in Figure~\ref{fig:graph_overview} has a {\Merge} set to 2. There is no merging in $e$, $i$ and $l$ because their {\ReachGrid}s contain only one grid. The rest are $b$, $c$ and $g$. In {\ReachGrid}($b$), for each grid in $L_1$, we make the {\Merge} checking. $G_{17}$ covers four grids $G_1$, $G_2$, $G_5$ and $G_6$ in $L_0$. In such four-grids region, only $G_2$ is reachable from $b$. The merging will not happen in $G_{17}$. It is the same case in $G_{18}$ and $G_{20}$. However, there are two grids, $G_9$ and $G_{14}$ covered by $G_{19}$ in $L_1$. As a result, the two grids in $L_0$ will be removed from {\ReachGrid}($b$) with $G_{19}$ being added instead. For the grid $G_{21}$ in $L_2$, the same checking in $L_1$ will be performed. Since, only $G_{19}$ is reachable, no merging happens. Finally, {\ReachGrid}($b$) = $\{2,19\}$. Similarly, we can have {\ReachGrid}($c$) = $\{12,14\}$ and {\ReachGrid}($g$) = $\{12,14\}$ where no merging occurs.

\begin{algorithm}[t]
	\caption{Maintain B-vertex}	
	\begin{scriptsize}
		\label{alg:Ini_B}
		\begin{algorithmic}[1]			
			\STATE {\bf Function} \textsc{Maintain-BVertex}(From-side vertex $v$, To-side vertex $v'$)
			\IF {{\GeoB}($v$) = \emph{true}}
				\STATE {{\bf return} \emph{false}}
			\ELSE
				\SWITCH {Type of $v'$}
				\CASE {B-vertex}
					\IF {{\GeoB}($v'$) = \emph{true}}
						\STATE {{\GeoB}($v$) $\leftarrow$ \emph{true}}
					\ELSIF {$v'.spatial$ $\neq$ NULL}
						\STATE {{\ReachGrid}($v$) $\leftarrow$ {\Grid}($v'.spatial$)}
					\ELSE
						\RETURN {\emph{false}}
					\ENDIF
				\ENDCASE
				\CASE {R-vertex}
					\STATE {{\RMBR}($v$) $\leftarrow$ {\MBR}({\RMBR}($v'$), $v'.spatial$)}
				\ENDCASE
				\CASE {G-vertex}
					\STATE {{\ReachGrid}($v$) $\leftarrow$ {\ReachGrid}($v'$)$\cup${\Grid}($v'.spatial$)}
				\ENDCASE
				\ENDSWITCH
			\ENDIF	
			\RETURN {\emph{true}}
		\end{algorithmic}
	\end{scriptsize}
\end{algorithm}

\subsection{SPA-Graph Maintenance}
\label{sec:maintenance}

When the structure of a graph is updated, i.e., adding or deleting edges and/or vertices, \GeoReach needs to maintain the SPA-Graph structure accordingly. Moreover, when the spatial attribute of a vertex changes, \GeoReach may need to maintain the \RMBR and/or \ReachGrid properties of that vertex and other connected vertices as well. As a matter of fact, all graph updates can be simulated as a combination of adding and/or deleting a set of edges. 

{\bf Adding an edge.} When an edge is added to the graph, the directly-influenced vertices are those that are connected to another vertex by the newly added edge. The spatial reachability information of the to-side vertex will not be influenced by the new edge. Based upon Lemmas~\ref{lemma:RMBR} and~\ref{lemma:ReachGrid}, the spatial reachability information, i.e., \RMBR or \ReachGrid, of the to-side vertex should be modified based on the the from-side vertex. On the other hand, the from-side vertex may remain the same or change. In the former case, there is no recursive updates required for the in-edge neighbors of the from-side vertex. Otherwise, the recursive updates are performed in the reverse direction until no change occurs or there is no more in-edge neighbor. A queue $Q$ will be exploited to track the updated vertices. When $Q$ is not empty, which means there are still some in-edge neighbors waiting for updates, the algorithm retrieves the next vertex in the queue. For such vertex, all its in-edge neighbors are updated by using the reachability information stored on this vertex. Updated neighbors will then be pushed into the queue. The algorithm halts when the queue is empty. Depending on category of the from-side vertex, corresponding maintenance functions, including {Maintain-BVertex}, {Maintain-RVertex} and {Maintain-GVertex} are used to update the newly added spatial reachability information.

Algorithm~\ref{alg:Ini_B} is used when the from-side vertex is a B-vertex. 
In algorithm~\ref{alg:Ini_B}, if the from-side vertex $v$ is already a B-vertex with {\GeoB}($v$) = \emph{true}. The added edge will never cause any change on $v$. Hence a false value is returned. In case {\GeoB}($v$) = \emph{false}, the algorithm considers type of the to-side vertex $v'$.

\begin{myitem}
	\item {\bf B-vertex.} If {\GeoB}($v'$) = \emph{true}, it is no doubt that {\GeoB}($v$) will be set to true and a true value will be returned. Otherwise, the algorithm checks whether $v'$ is spatial. If it is, {\ReachGrid}($v$) is updated with {\Grid}($v'spatial$). Otherwise, the algorithm returns false because $v$ is not changed.
	
	\item {\bf R-vertex.} In such case, it is certain that $v$ will be updated to an R-vertex. The algorithm merely updates {\RMBR}($v$) with {\MBR}({\RMBR}($v'$), $v'.spatial$).
	
	\item {\bf G-vertex.} It is similar to the R-vertex case. Type of $v'$ can decide that $v$ should be a G-vertex and the algorithm updates {\ReachGrid}($v$) with {\ReachGrid}($v'$)$\cup${\Grid}($v'.spatial$)
\end{myitem}

{Maintain-BVertex} and {Maintain-RVertex} are what we use in the initialization. However, there is a new condition that should be taken into consideration. When the from-side vertex $v$ is an R-vertex and the to-side vertex $v'$ is a G-vertex, the algorithm needs to update the {\RMBR}($v$) with {\ReachGrid}($v'$). Under such circumstance, first a dummy {\RMBR}($v'$) will be constructed using {\ReachGrid}($v'$). Although it is not the exact {\RMBR} of $v'$, it is still precise. Error of the width and height will not be greater than size of a grid cell. 
No matter what function is invoked to update the from-side vertex, {\GeoReach} takes into account the system parameters {\MR} and {\MG} are checked on {\RMBR} and {\ReachGrid}, respectively.


%
%

{\bf Deleting an edge.}  When an edge is removed, the to-side vertex will be not impacted by the deleting which is the same with adding an edge. To maintain the correctness of spatial reachability information stored on the from-side vertex, the only way is to reinitialize its spatial reachability information according to all its current out-edge neighbors. If its structure is different from the original state due to the deleting, the structure of all its in-edge neighbors will be rebuilt recursively. A queue $Q$ is used to keep track of the changed vertices. The way \GeoReach maintains the queue and the operations on each vertex in the queue are similar to the AddEdge procedure. Maintenance cost of deleting an edge will be $O(kn^3)$ because the whole {\GeoReach} index may be  reinitialized.


\newcommand{\uniprotc}{uniprot150m}
\newcommand{\uniprota}{uniprot22m}
\newcommand{\uniprotb}{uniprot100m}
\newcommand{\patent}{patent}
\newcommand{\gouniprot}{go-uniprot}
\newcommand{\citeseerx}{citeseerx}

\section{Experimental Evaluation}
\label{sec:experiment}

In this section, we present a comprehensive experimental evaluation of {\GeoReach} performance. We compare the following approaches: {\GeoF}, {\GeoMa}, {\GeoMb}, {\GeoP}, {\GeoRMBR} and {\SpaReach}. {\GeoF}, {\GeoMa} and {\GeoMb} are approaches that store only {\ReachGrid} by setting {\MG} to the total number of grids in the space and {\MR} to A where A represent the area of the whole 2D space. Their difference lies in the value of {\Merge}. {\GeoF} is an approach where {\Merge} is 0. In such approach, no higher layer grids are merged. {\Merge} is set to 2 and 3 respectively in {\GeoMa} and {\GeoMb}. {\GeoP} is an approach in which {\Merge} = 0, {\MG} = 200 and {\MR} = A. In such approach, reachable grids in {\ReachGrid} will not be merged. If the number of reachable grids of {\ReachGrid}($v$) is larger than 200 then $v$ will be degraded to an R-vertex. Since {\MR} = A, there will be no B-vertex. In {\GeoRMBR}, {\MG} = 0, {\MR} = A, hence only {\RMBR} s are stored. In all {\ReachGrid} related approaches, the total space is split into $128\times 128$ pieces in the highest grid layer. {\SpaReach} approach is implemented with both spatial index and reachability index. Graph structure is stored in Neo4j graph database. Reachability index is stored as attributes of each graph vertex in Neo4j database. Reachability index we use is proposed in~\cite{YAI+2013}. Spatial index used {\SpaReach} approaches is implemented by gist index in postgresql. To integrate Neo4j and postgresql databases, for each vertex in the graph, we assign it an id to uniquely identify it.

{\bf Experimental Environment.} The source code for evaluating query response time is implemented in Java and compiled with java-7-openjdk-amd64. Source codes of index construction are implemented in c++ and complied using g++ 4.8.4. Gist index is constructed automatically by using command line in Postgresql shell. All evaluation experiments are run on a computer with an 3.60GHz CPU, 32GB RAM running Ubuntu 14.04 Linux OS.


\begin{table}[t]
	\small
	\centering
	\caption{\small \label{real_datasets} Graph Datasets (K = $10^3$)}
	\begin{tabular}{|l|c|c|c|c|}
		\hline {\bf Dataset} &$|V|$&$|E|$&$d_{avg}$&$l$\\
		\hline citeseerx~~&6540K&15011K&2.30&59\\
		\hline go-uniprot~~&6968K&34770K&4.99&21\\
		\hline patent&3775K~~&16519K&4.38&32\\
		\hline uniprot22m~~&1595K&1595K&1.00&4\\
		\hline uniprot100m~~&16087K&16087K&1.00&9\\
		\hline uniprot150m~~&25038K&25038K&1.00&10\\
		\hline		
	\end{tabular}
\end{table}

{\bf Datasets.} We evaluate the performance of our methods using six real datasets~\cite{CHS+2013,YAI+2013} (see Table~\ref{real_datasets}). Number of vertices and edges are listed in column $|V|$ and $|E|$. Column $d_{avg}$ and $l$ are average degree of vertices and length of the longest path in the graph, respectively. Citeseerx and patent are real life citation graphs extracted from CiteSeerx\footnote{\scriptsize http://citeseer.ist.psu.edu/} and US patents\footnote{\scriptsize http://snap.stanford.edu/data/}~\cite{YAI+2013}. Go-uniprot is a graph generated from Gene Ontology and annotation files from Uniprot\footnote{\scriptsize http://www.uniprot.org/}~\cite{YAI+2013}. Uniprot22m, uniprot100m and uniprot150m are RDF graphs from UniProt database~\cite{YAI+2013}. The aforementioned datasets represent graphs that possess no spatial attributes. For each graph, we simulate spatial data by assigning a spatial location to a subset of the graph vertices. During the experiments, we change the ratio of spatial vertices to the total number of vertices from 20\% to 80\%. During the experiments, we vary the spatial distribution to be: uniform, zipf, and clustered distributions. Unless mentioned otherwise, the number of spatial clusters is set to 4 by default.

\subsection{Query Response Time}

In this section, we fist compare the query response time performance of {\SpaReach} to our {\GeoP} approach. Afterwards, we change tunable parameters in {\GeoReach} to evaluate influence of these thresholds. For each dataset, we change the spatial selectivity of the input query rectangle from 0.0001 to 0.1. For each query spatial selectivity, we randomly generate 500 queries by randomly selecting 500 random vertices and 500 random spatial locations of the query rectangle. The reported query response time is calculated as the average time taken to answer the 500 queries.

\begin{figure*}	
	\centering
	\begin{subfigure}[t]{0.23\textwidth}
		\centering
		\includegraphics[width=1.0\linewidth]{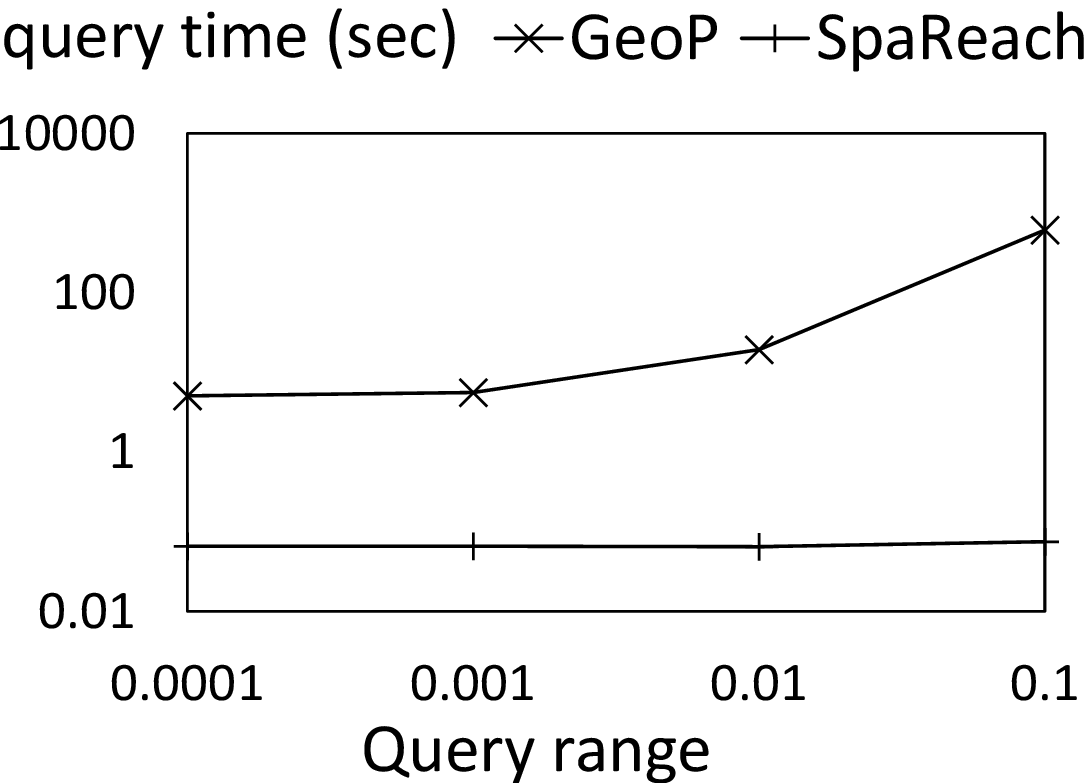}
		\caption{\uniprotc}
		\label{fig:query_time_uniprot150m}
	\end{subfigure}
	~
	\begin{subfigure}[t]{0.23\textwidth}
		\centering
		\includegraphics[width=1.0\linewidth]{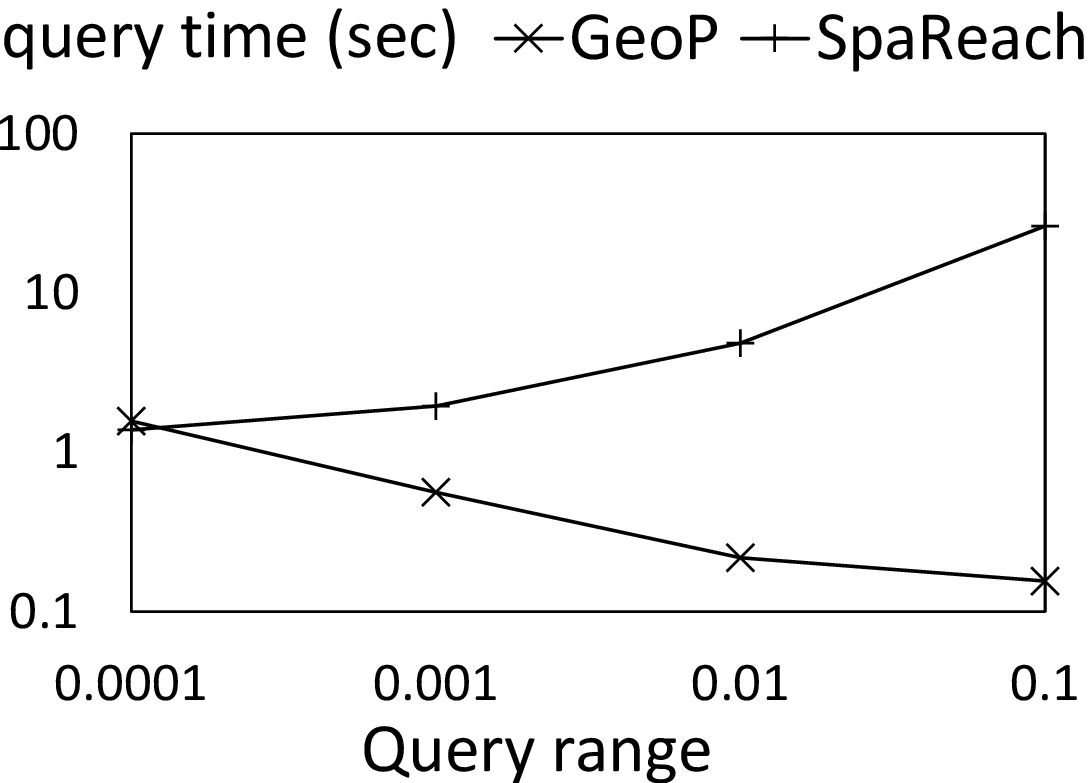}
		\caption{\patent}
		\label{fig:query_time_Patents}
	\end{subfigure}
	~
	\begin{subfigure}[t]{0.23\textwidth}
		\centering
		\includegraphics[width=1.0\linewidth]{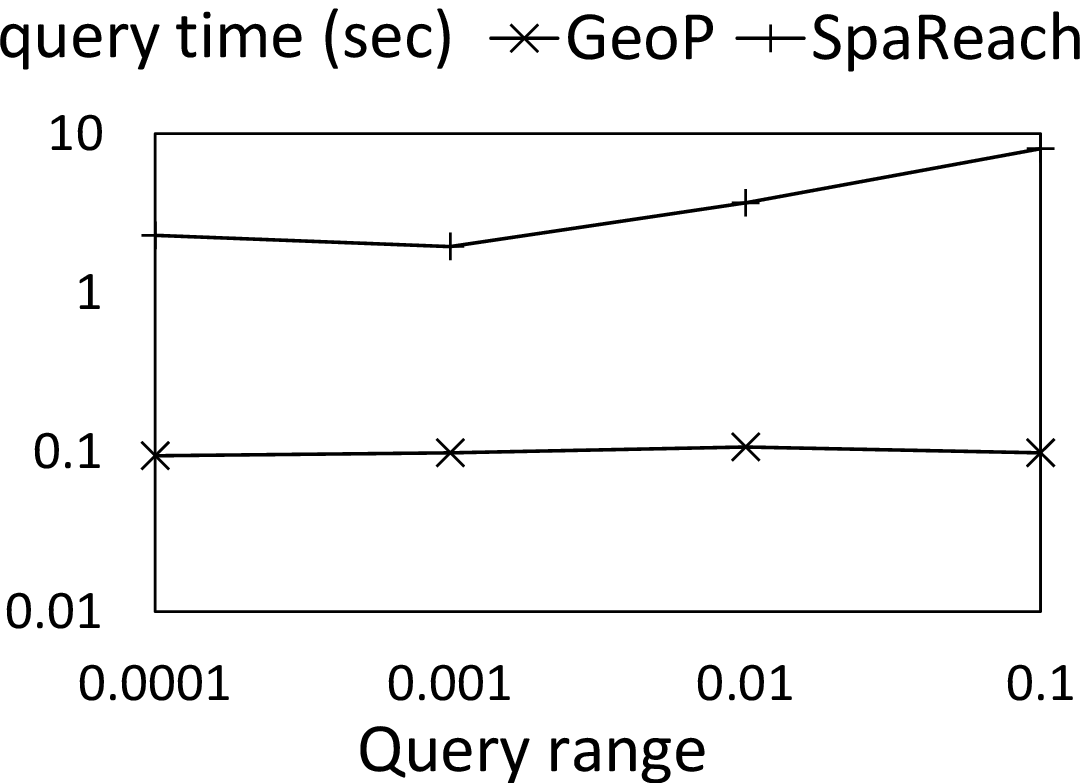}
		\caption{\gouniprot}
		\label{fig:query_time_go_uniprot}
	\end{subfigure}
	~
	\begin{subfigure}[t]{0.23\textwidth}
		\centering
		\includegraphics[width=1.0\linewidth]{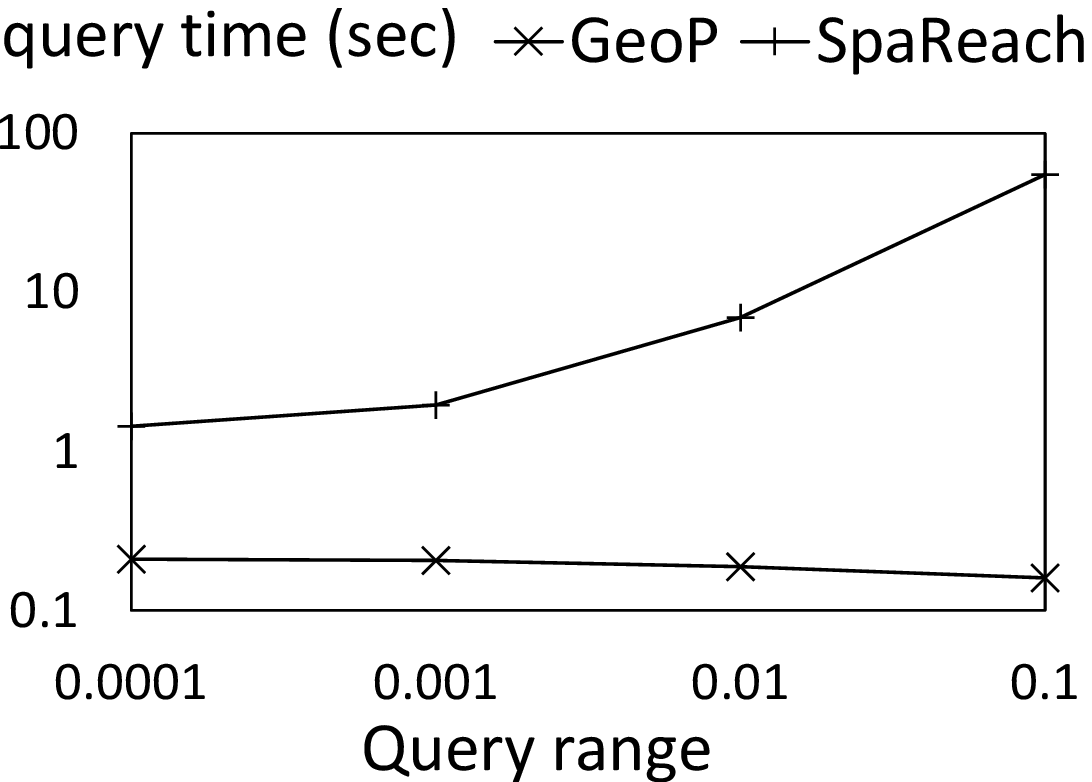}
		\caption{\citeseerx}
		\label{fig:query_time_citeseerx}
	\end{subfigure}
	\caption{\small Query response time (80\% spatial vertex ratio, randomly-distributed spatial data, and spatial selectivity ranging from 0.0001 to 0.1)}
	\label{fig:query_time}	
\end{figure*}

Figure~\ref{fig:query_time} depicts the query response time of {\GeoP} and {\SpaReach} on four datasets. 80\% of vertices in the graph are spatial and they are randomly-distributed in space. For brevity, we omit the results of the other two datasets, i.e., {\uniprota} and {\uniprotb}, since they have almost the same graph structure and exihibit the same performance. As it turns out In Figure~\ref{fig:query_time}, {\GeoP} outperforms {\SpaReach} for any query spatial selectivity in {\uniprotc}, {\gouniprot} and {\citeseerx}. For these datasets, {\SpaReach} approach cost more time when query selectivity increases. When we increasing the query range size, the range query step tends to return a larger number of spatial vertices. Hence, the graph reachability checking step has to check more spatial vertices. Figure~\ref{fig:query_time_go_uniprot} and~\ref{fig:query_time_citeseerx} show similar experiment results. In conclusion, {\GeoP} is much more query-efficient in relatively sparse graphs. Patent dataset is the densest graph with richer reachability information. Figure~\ref{fig:query_time_Patents} indicates that even when spatial selectivity set to 0.0001, {\GeoP} can achieve almost the same performance as {\SpaReach}. When spatial selectivity increases, {\GeoP} outperforms {\SpaReach} again. In a denser graph, the performance difference between the two approaches is smaller than in sparse graphs especially when the spatial selectivity is low. 

\begin{table*}[t]
	\small
	\captionsetup{font = small}
	\caption{\small \label{query_time_ours} Query Response Time in three datasets, 80\% spatial vertex ratio, and spatial selectivity ranging from 0.0001 to 0.1}
	\centering	
	\begin{tabular}{|l|c|c|c|c|c|c|c|c|c|c|c|c|c|c|c|}
		\hline
		&\multicolumn{5}{|c|}{uniprot150m}&\multicolumn{5}{|c|}{patent}&\multicolumn{5}{|c|}{citeseerx}\\
		\hline Selectivity&MT0&MT2&MT3&GeoP&RMBR&MT0&MT2&MT3&GeoP&RMBR&MT0&MT2&MT3&GeoP&RMBR\\
		\hline 0.0001&68&68&67&66&66&643&762&741&1570&2991&202&212&203&210&234\\
		\hline 0.001&65&77&78&66&65&168&258&185&559&1965&34&460&471&207&215\\
		\hline 0.01&66&66&65&65&65&87&143&98&217&915&32&408&410&189&200\\
		\hline 0.1&69&65&65&75&66&51&108&59&155&348&33&399&399&160&183\\
		\hline
	\end{tabular}
\end{table*}

Table~\ref{query_time_ours} compares the query response time of all our approaches for the {\uniprotc}, {\patent} {\gouniprot} and {\citeseerx} datasets with randomly distributed spatial vertices and spatial ratio of 80\%. In {\uniprotc}, all our approaches almost have the same performance. The same pattern happens with the {\uniprota}, {\uniprotb} and {\gouniprot} datasets. So we use {\uniprotc} as a representative. 


For the {\patent} graph with random-distributed spatial vertices and spatial ratio of 20\%, query efficiency difference can be easily caught. {\GeoF} keeps information of exact reachable grids of every vertex which brings us fast query speed, but also the highest storage overhead. {\RMBR} stores general spatial boundary of reachable vertices which is the most scalable. However, such approach spend the most time in answering the query. Since {\GeoMb} is an approach that {\Merge} is set to 3, just few grids in {\GeoMb} are merged. As a result, its query time is merely little bit longer than {\GeoF}. There are more grids getting merged in {\GeoMa} than in {\GeoMb}. Inaccuracy caused by more integration lowers efficiency of {\GeoMb} in query. {\GeoP} is combination of {\ReachGrid} and {\RMBR}. Its query efficiency is lower than {\GeoF} and better than {\GeoRMBR}. In this case, {\GeoMa} outperforms {\GeoP}. But it is not always the case. By tuning {\MG} to a larger number, {\GeoP} can be more efficient in query.

In {\citeseerx}, {\GeoF} keeps the best performance as expected. Performance of {\GeoP} is in between {\GeoF} and {\GeoRMBR} as what is shown in \patent. But {\GeoMa} and {\GeoMb} reveal almost the same efficiency and they are worse than {\GeoRMBR}. Distinct polarized graph structure accounts for the abnormal appearance. In {\citeseerx}, all vertices can be divided into two groups. One group consists of vertices that cannot reach any vertex. The other group contains a what we call center vertex. The center vertex has huge number of out-edge neighbor vertices and is connected by huge number of vertices as well. Because the center vertex can reach that many vertices, it can reach nearly all grid cells in space. As a result, vertices that can reach the center vertex can also reach all grid cells in space. So no matter what value is {\MG}, reachable grids in {\ReachGrid} of these vertices will be merged into only one grid in a lower layer until to the bottom layer which is the whole space. Then such {\ReachGrid} can merely function as a {\GeoB} which owns poorer locality than {\RMBR}.

\subsection{Storage Overhead}

\begin{figure*}[t]	
	\centering
	\begin{subfigure}[t]{0.23\textwidth}
		\centering
		\includegraphics[width=1.0\linewidth]{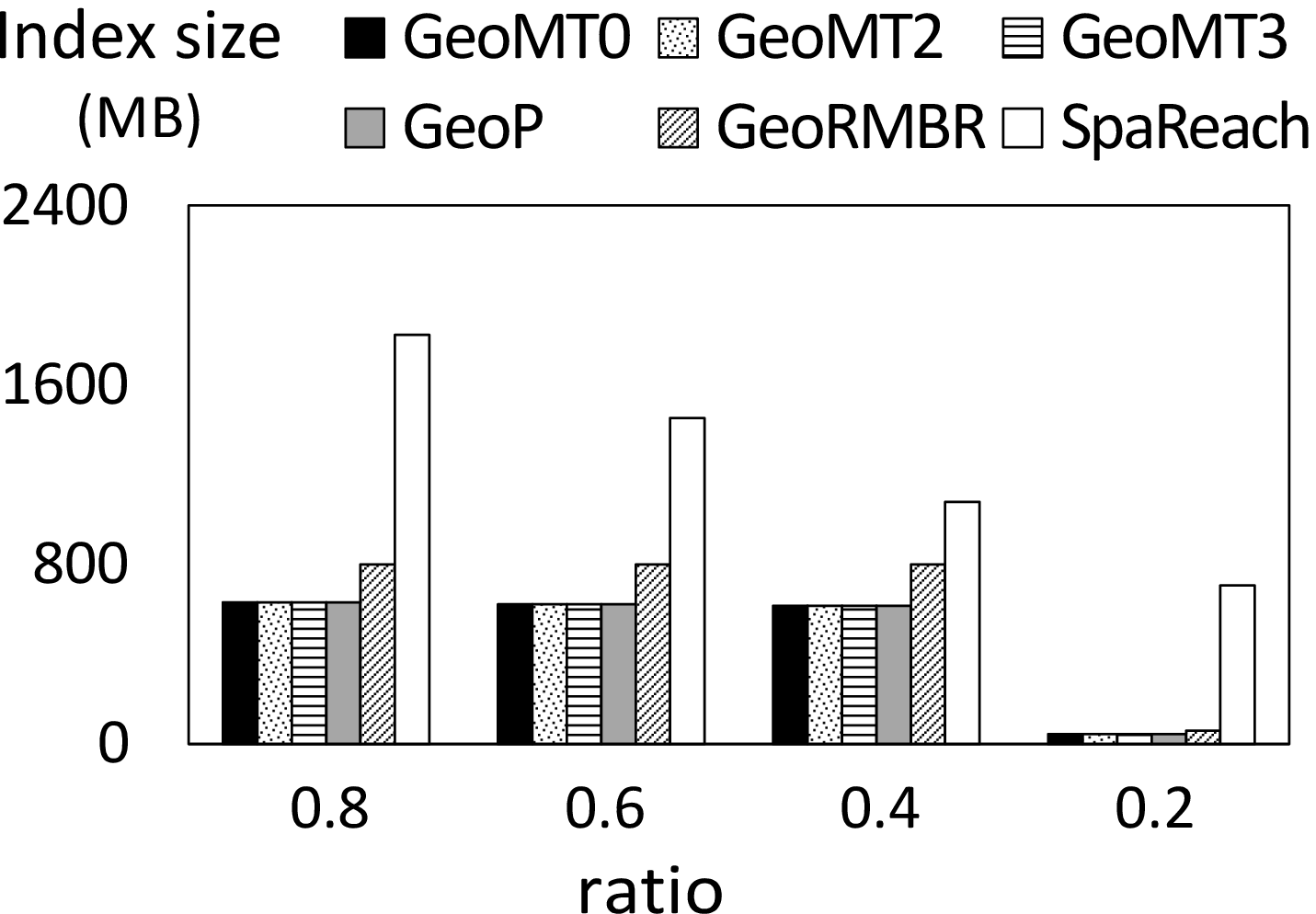}
		\caption{uniprot150m}
		\label{fig:index_size_uni150}
	\end{subfigure}
	~
	\begin{subfigure}[t]{0.23\textwidth}
		\centering
		\includegraphics[width=1.0\linewidth]{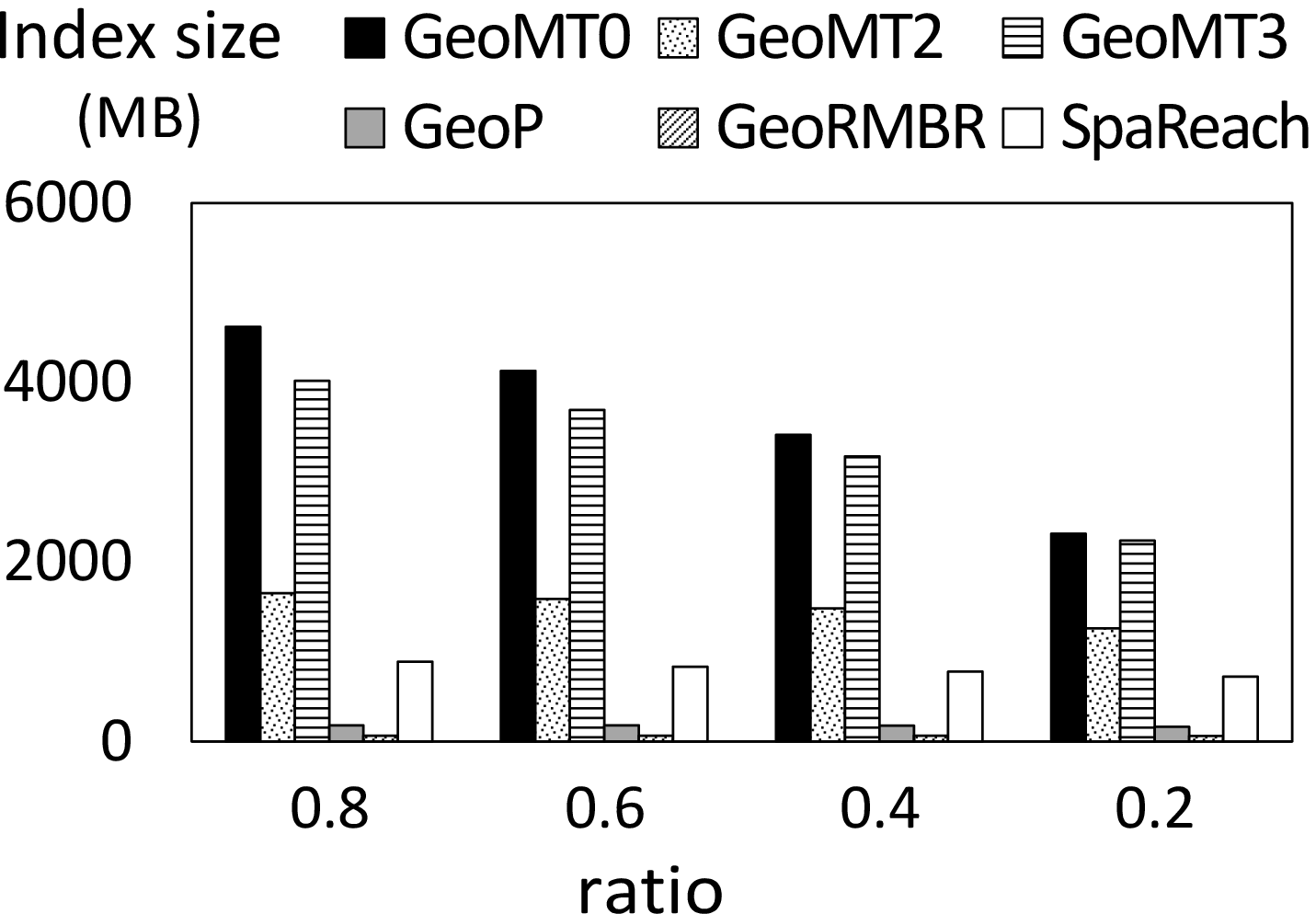}
		\caption{patent}
		\label{fig:index_size_patent_random}
	\end{subfigure}
	~
	\begin{subfigure}[t]{0.23\textwidth}
		\centering
		\includegraphics[width=1.0\linewidth]{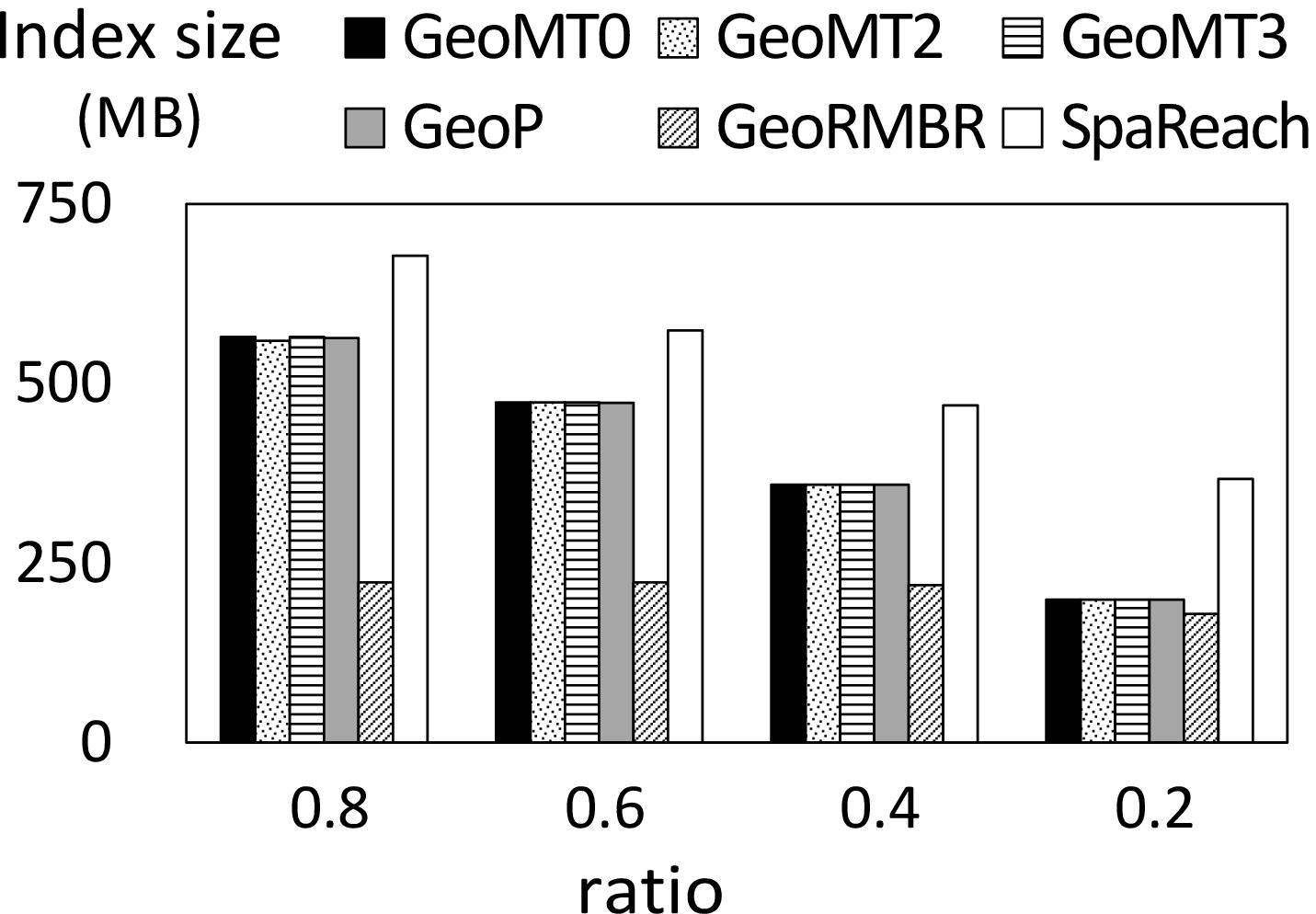}
		\caption{go-uniprot}
		\label{fig:index_size_go_uniprot}
	\end{subfigure}
	~
	\begin{subfigure}[t]{0.23\textwidth}
		\centering
		\includegraphics[width=1.0\linewidth]{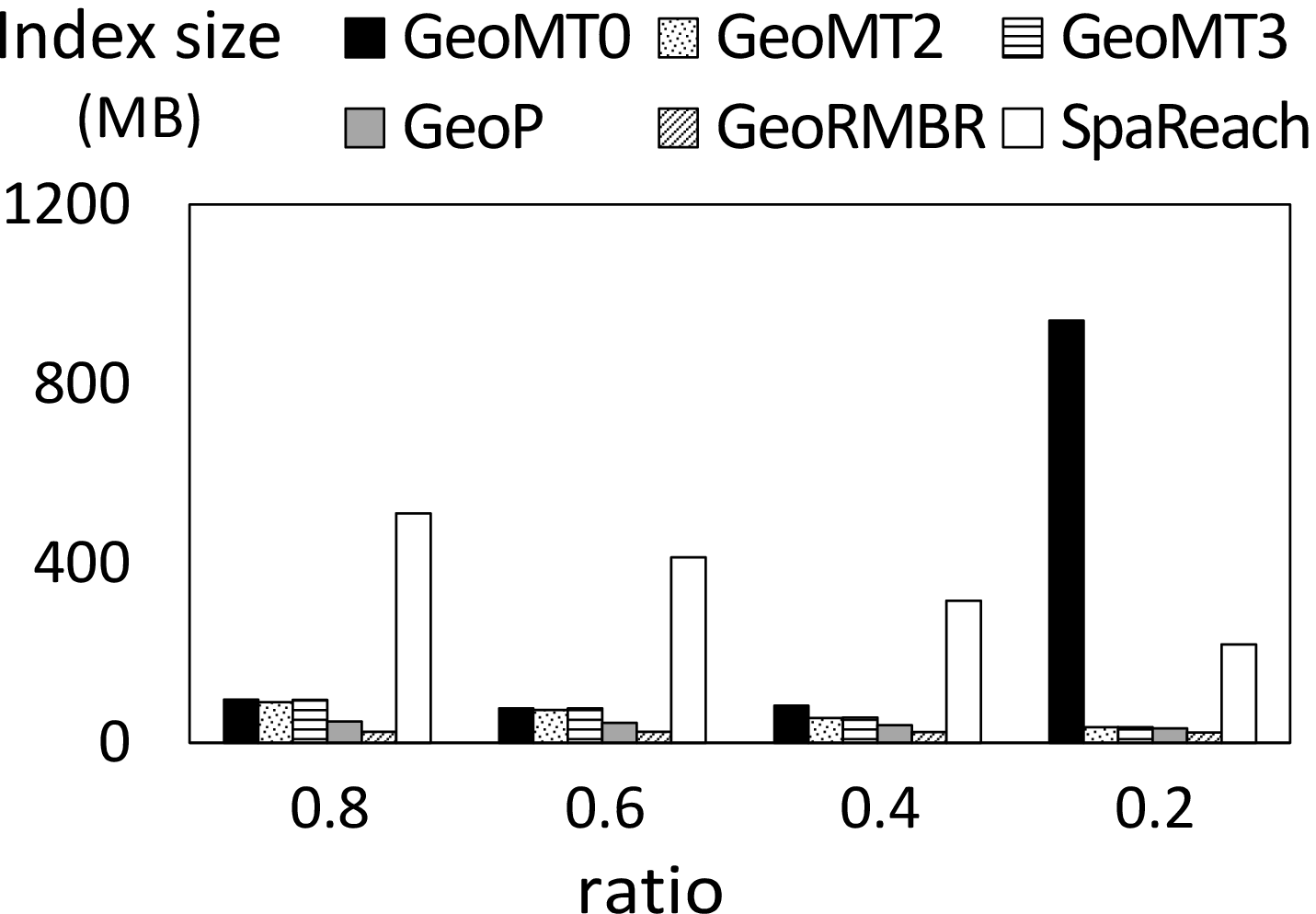}
		\caption{citeseerx}
		\label{fig:index_size_citeseerx}
	\end{subfigure}
	\caption{\small Storage Overhead (Randomly distributed, spatial vertex ratio from 0.8 to 0.2)}
	\label{fig:index_size}
\end{figure*}

\begin{figure*}[t]
	\centering
	\begin{subfigure}[t]{0.23\textwidth}
		\centering
		\includegraphics[width=1.0\linewidth]{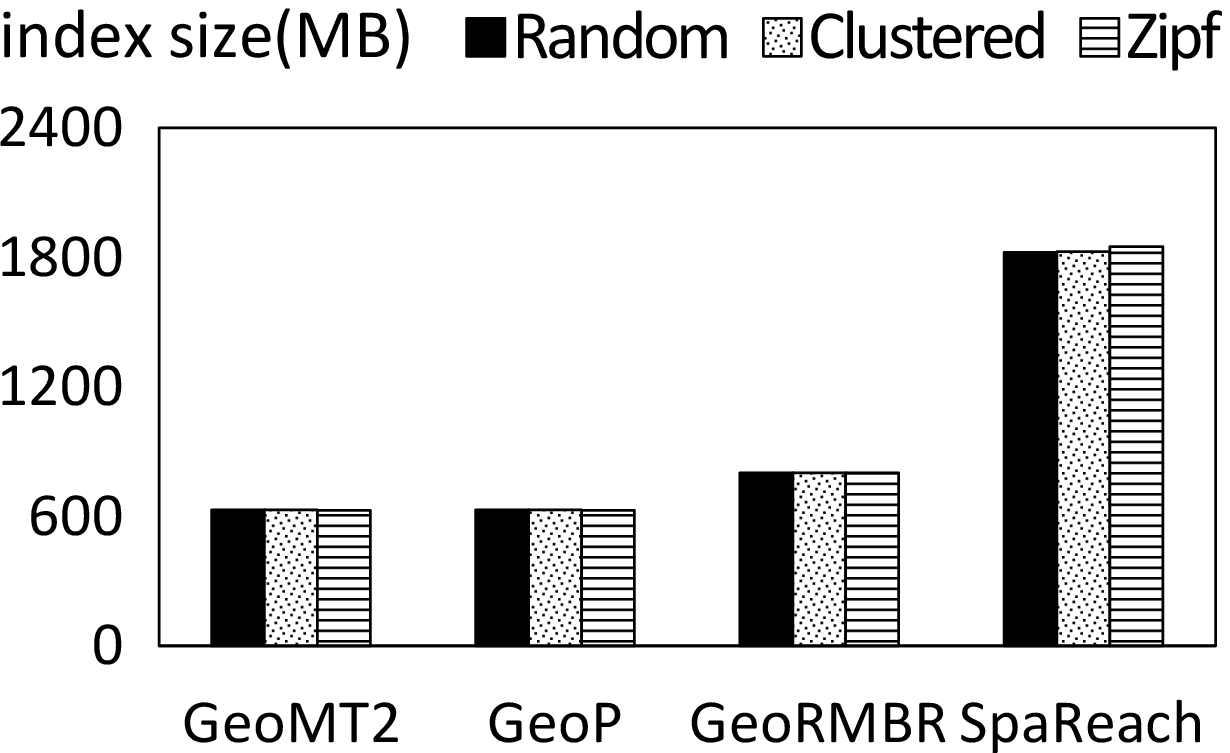}
		\caption{uniprot150m}
		\label{fig:index_size_uni150_vary_distribution}
	\end{subfigure}
	~
	\begin{subfigure}[t]{0.23\textwidth}
		\centering
		\includegraphics[width=1.0\linewidth]{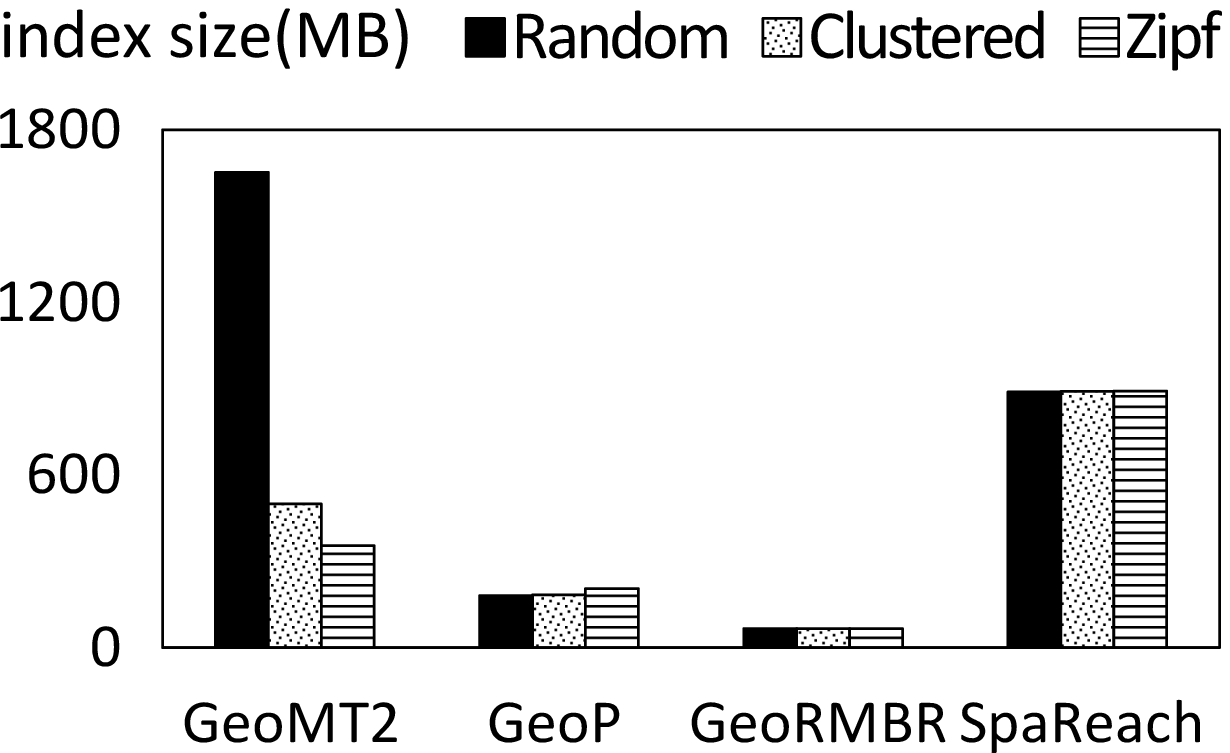}
		\caption{patent}
		\label{fig:index_size_patent_vary_distribution}
	\end{subfigure}
	~
	\begin{subfigure}[t]{0.23\textwidth}
		\centering
		\includegraphics[width=1.0\linewidth]{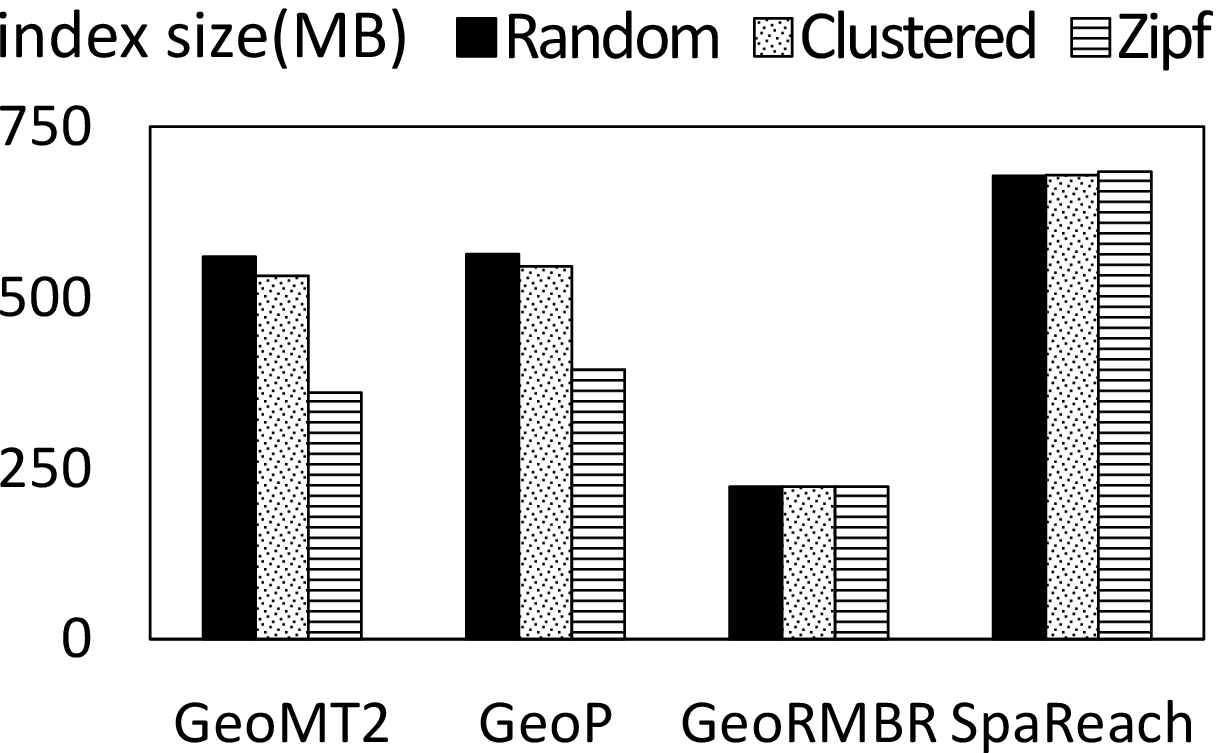}
		\caption{go-uniprot}
		\label{fig:index_size_go_uniprot_vary_distribution}
	\end{subfigure}
	~
	\begin{subfigure}[t]{0.23\textwidth}
		\centering
		\includegraphics[width=1.0\linewidth]{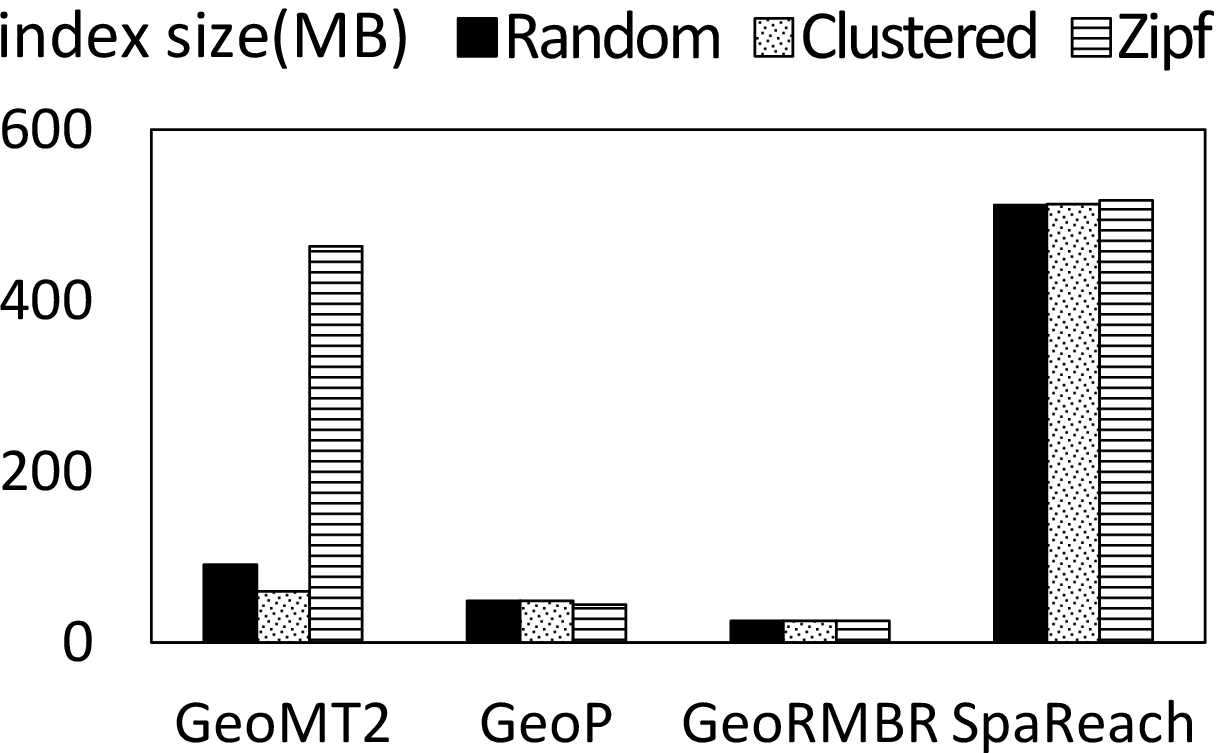}
		\caption{citeseerx}
		\label{fig:index_size_citeseerx_vary_distribution}
	\end{subfigure}
	\caption{\small Storage Overhead for varying spatial data distribution (randomly, cluster and zipf distributed) and 0.8 spatial vertex ratio)}
	\label{fig:index_size_vary_distribution}
\end{figure*}


Figure~\ref{fig:index_size_uni150} gives the storage overhead of all approaches for the {\uniprotc} dataset. In this experiment, the spatial vertices are randomly distributed in space. Since {\uniprota} and {\uniprotb} share the same pattern with {\uniprotc} (even spatial distribution of vertices varies), they are not shown in the figure. The experiments show that {\GeoReach} and all its variants require less storage overhead than {\SpaReach} because of the additional overhead introduced by the spatial index. When there are less spatial vertices, {\SpaReach} obviously occupies less space because size of spatial index lessens. However, {\SpaReach} always requires more storage than any other approaches. Storage overhead of {\GeoReach} approaches shows a two-stages pattern which means it is either very high (ratio = 0.8, 0.6 and 0.4) or very low (ratio = 0.2). The reason is as follows. These graphs are sparse and almost all vertices reach the same vertex. This vertex cannot reach any other vertex. Let us call it an end vertex. If the end vertex is a spatial vertex, then all vertices that can reach the end vertex will keep their spatial reachability information (no matter what category they are) in storage. But if it is not, majority of vertices will store nothing for spatial reachability information. {\GeoF} and {\GeoP} are of almost the same index size because of sparsity and end-point phenomenon in these graphs. Such characteristic causes that almost each vertex can just reach only one grid which makes {\MG} invalid in approach {\GeoF} (number of reachable grids is always less than {\MG}) which makes {\GeoF} and {\GeoP} have nearly the same size. For similar reason, {\Merge} becomes invalid in these datasets which makes {\GeoMa} and {\GeoMb} share the same index size with {\GeoF} and {\GeoP}. We also find out that index size of {\GeoRMBR} is slightly larger than {\GeoF} approaches. Intuitively, {\RMBR} should be more scalable than {\ReachGrid}. But most of the vertices in these three graphs can reach only one grid. In {\GeoRMBR}, for each vertex that have reachable spatial vertices, we assign an {\RMBR} which will be stored as coordinates of {\RMBR}'s top-left and lower-right points. It is more scalable to store one grid id than two coordinates. So when a graph is highly sparse, index size of {\GeoF} is possible to be less than {\GeoRMBR}.

Figure~\ref{fig:index_size_go_uniprot} shows that in {\gouniprot} all {\GeoReach} approaches performs better than {\SpaReach}. When we compare all the {\GeoReach} approaches, {\GeoF}, {\GeoMa} and {\GeoMb} lead to almost the same storage overhead. That happens due to the fact that {\gouniprot} is a very sparse graph. A vertex can only reach few grids in the whole space. Grid cells in {\ReachGrid} can hardly be spatially adjacent to each other which causes no integration. The graph sparsity makes the number of reachable grids in {\ReachGrid} always less than {\MG} which leads to less R-vertices and more G-vertices. In consequence, {\gouniprot}, {\GeoF}, {\GeoMa}, {\GeoMb} and {\GeoP} lead to the same storage overhead. It is rational that {\GeoRMBR} requires the least storage because {\RMBR} occupies less storage than {\ReachGrid}.

When graphs are denser, results become more complex. Figure~\ref{fig:index_size_patent_random} shows index size of different approaches in patent dataset with randomly-distributed spatial vertices. {\GeoRMBR} and {\GeoP}, take the first and the second least storage and are far less than other approaches because both of them use {\RMBR} which is more scalable. {\GeoF} takes the most storage in all spatial ratios for that {\ReachGrid} takes high storage overhead. {\GeoMa} and {\GeoMb} require less storage than {\GeoF} because spatially-adjacent reachable grids in {\GeoF} are merged which brings us scalability. {\GeoMb} are more scalable than {\GeoMa} because {\Merge} in {\GeoMa} is 2 which causes more integration. There are three approaches, {\GeoMb}, {\GeoP} and {\GeoRMBR}, that outperform {\SpaReach} approach. By tuning parameters in {\GeoReach}, we are able achieve different performance in storage overhead and can also outperform {\SpaReach}.

Figure~\ref{fig:index_size_citeseerx} depicts index size of all approaches in {\citeseerx} with randomly distributed spatial vertices. Spatial vertices ratio ranges from 0.8 to 0.2. All {\GeoReach} approaches outperform {\SpaReach} except for one outlier when spatial vertices ratio is 0.2. {\GeoF} consumes huge storage. This is caused by the center vertex which is above-mentioned. Recall that large proportion of {\ReachGrid} contains almost all grids in space. After bitmap compression, it will cause low storage overhead. This is why when spatial vertices ratio is 0.8, 0.6 and 0.4, {\GeoF} consumes small size of index. When the ratio is 0.2, there are less spatial vertices. Although graph structure does not change, the center vertex reach less spatial vertices and less grids. Then the bitmap compression brings no advantage in storage overhead.

Figure~\ref{fig:index_size_vary_distribution} shows the impact of spatial data distribution on the storage cost. {\GeoF}, {\GeoMa} and {\GeoMb} are all {\ReachGrid}-based approaches. Spatial data distribution of vertices influences all approaches the same way. For all datasets, {\SpaReach} is not influenced by the spatial data distribution. {\SpaReach} consists of two sections: (1)~The reachability index size is determined by graph structure and (2)~The spatial index size is directly determined by number of spatial vertices. Hence, {\SpaReach} exhibits the same storage overhead for different spatial data distributions. When spatial vertices distribution varies, {\GeoRMBR} also keeps stable storage overhead. This is due to the fact that the storage overhead for each {\RMBR} is a constant and the number of stored {\RMBR} s is determined by the graph structure and spatial vertices ratio, and not by the spatial vertices distribution. Spatial data distribution can only influence the shape of each {\RMBR}.

Figure~\ref{fig:index_size_uni150_vary_distribution} shows that each approach in {\GeoReach} keeps the same storage overhead under different distributions in {\uniprotc}. As mentioned before, {\GeoF}, {\GeoMa}, {\GeoMb} and {\GeoP} actually represent the same data structure since there is only a single reachable grid in {\ReachGrid}. When there is only one grid reachable, varying  the spatial distribution becomes invalid for all approaches which use {\ReachGrid}.

Figure~\ref{fig:index_size_patent_vary_distribution} and~\ref{fig:index_size_go_uniprot_vary_distribution} shows that the storage overhead introduced by {\ReachGrid}-based approaches decreases when spatial vertices become more congested. Randomly distributed spatial data is the least congested while zipf distributed is the most. The number of reachable spatial vertices from each vertex do not change but these reachable spatial vertices become more concentrated in space. This leads to less reachable grids in {\ReachGrid}. 

Figure~\ref{fig:index_size_citeseerx_vary_distribution} shows that when spatial vertices are more congested, {\ReachGrid} based approaches, i.e., {\GeoF}, {\GeoMa} and {\GeoMb}, tend to be less scalable. Recall that {\citeseerx} dataset is a polarized graph with a center vertex. One group contains vertices that can reach huge number of vertices (about 200,000) due to the center vertex. When spatial vertices are more concentrated and that will lead to more storage overhead. 


\begin{figure*}[t]	
	\centering
	\begin{subfigure}[t]{0.23\textwidth}
		\centering
		\includegraphics[width=1.0\linewidth]{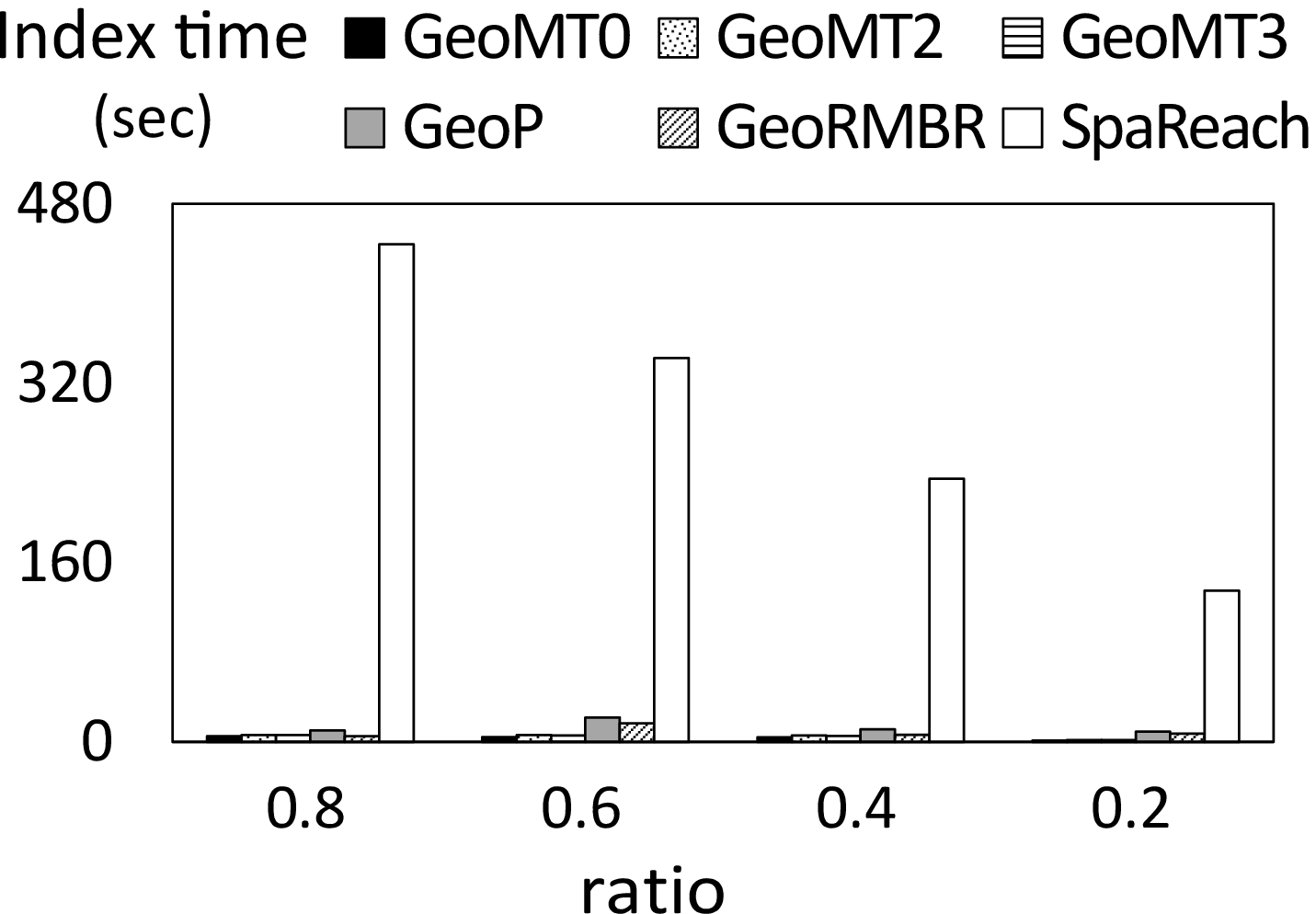}
		\caption{uniprot150}
		\label{fig:ini_time_uniprotenc_150}
	\end{subfigure}
	~
	\begin{subfigure}[t]{0.23\textwidth}
		\centering
		\includegraphics[width=1.0\linewidth]{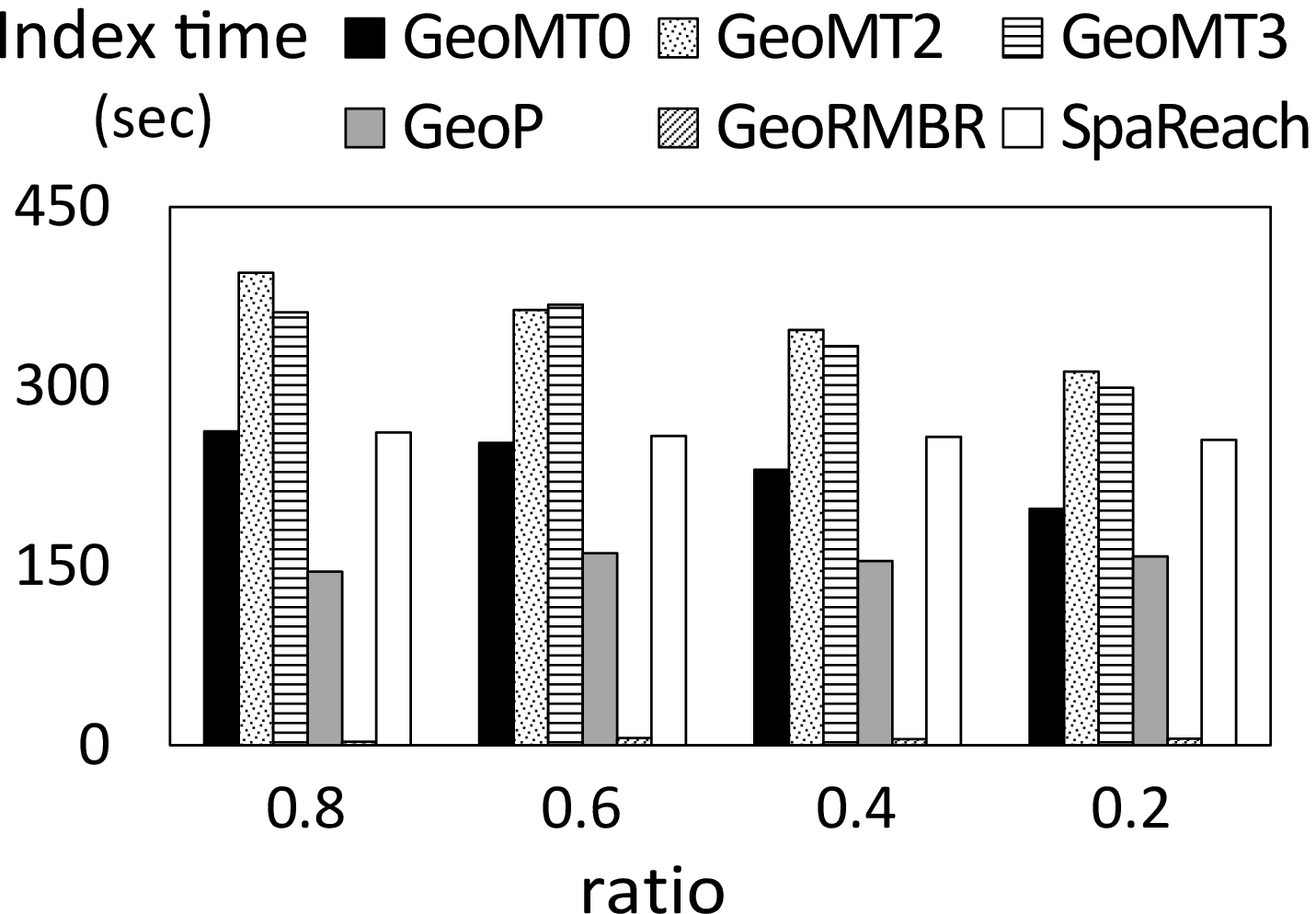}
		\caption{patent}
		\label{fig:ini_time_Patent}
	\end{subfigure}
	~
	\begin{subfigure}[t]{0.23\textwidth}
		\centering
		\includegraphics[width=1.0\linewidth]{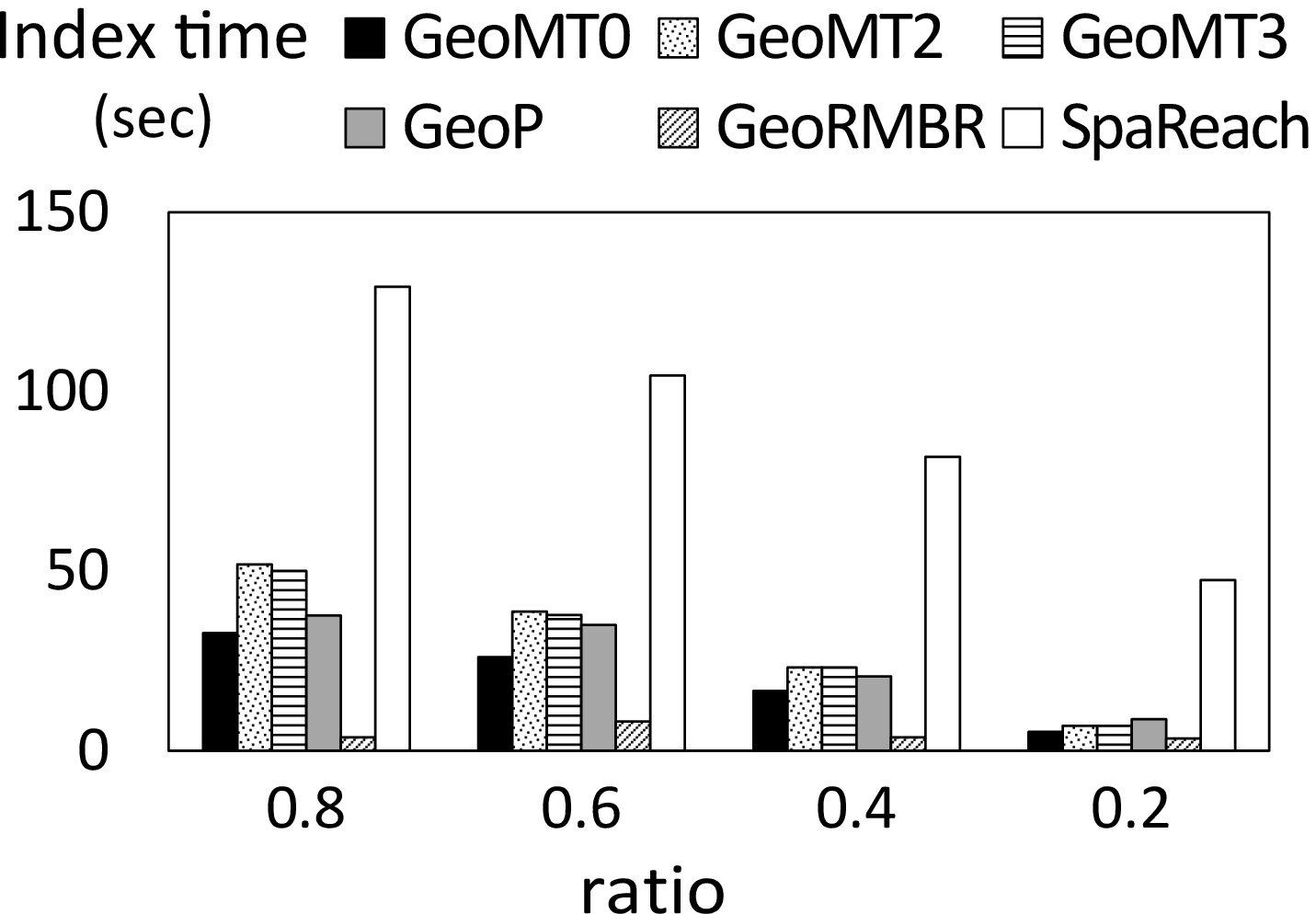}
		\caption{go-uniprot}
		\label{fig:ini_time_go_uniprot}
	\end{subfigure}
	\begin{subfigure}[t]{0.23\textwidth}
		\centering
		\includegraphics[width=1.0\linewidth]{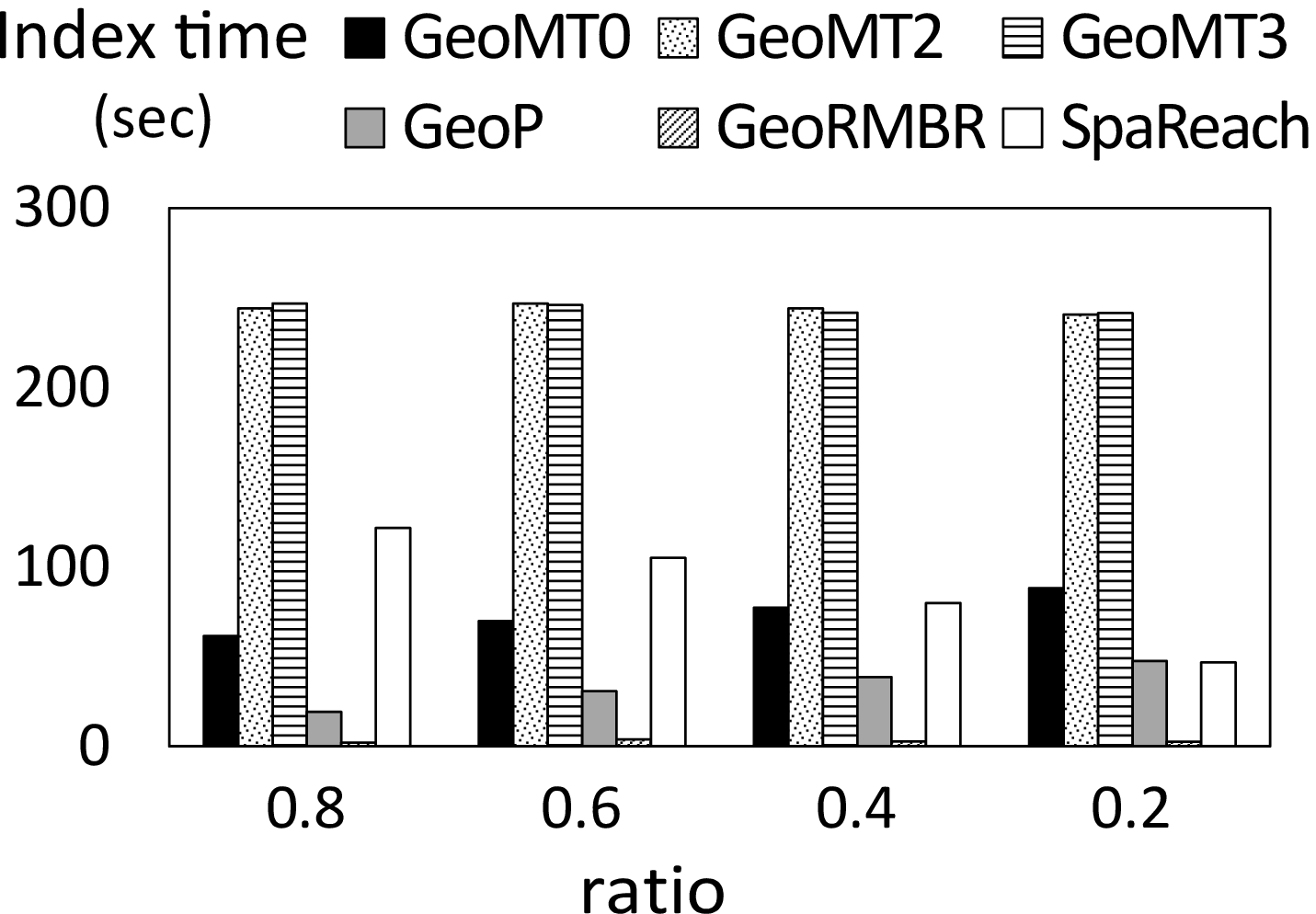}
		\caption{citeseerx}
		\label{fig:ini_time_citeseerx}
	\end{subfigure}
	\caption{\small Initialization time (Randomly distributed, spatial vertex ratio from 0.8 to 0.2)}\label{fig:ini_time}	
\end{figure*}

\subsection{Initialization time}

In this section, we evaluate the index initialization time for all considered approaches. For brevity, we only show the performance results for four datasets, {\uniprotc}, {\patent}, {\gouniprot} and {\citeseerx}, since {\uniprota}, {\uniprotb} and {\uniprotc} datasets exhibit the same performance. Figure~\ref{fig:ini_time_uniprotenc_150} shows that {\SpaReach} requires much more construction time than the other approaches under all spatial ratios. Although these graphs are sparse, they contain large number of vertices. This characteristic causes huge overhead in constructing a spatial index which dominates the initialization time in {\SpaReach}. Hence, {\SpaReach} takes much more time than all other approaches. However, the {\SpaReach} initialization time decreases when decreasing the number spatial vertices since the spatial index building step deals with less spatial vertices in such case. However, {\SpaReach} remains the worst even when the spatial vertex ratio is set to 20\%.

Figures~\ref{fig:ini_time_Patent} and~\ref{fig:ini_time_citeseerx} gives the initialization time for both the {\patent} and {\citeseerx} datasets, respectively. {\GeoRMBR} takes significantly less initialization time compared to all other approaches. {\GeoP} takes less time than the rest of approaches because it is {\ReachGrid} of partial vertices whose number of reachable grids are less than {\MG} that are calculated. In most cases, {\GeoF} can achieve almost equal or better performance compared to {\SpaReach} while {\GeoMa} and {\GeoMb} requires more time due to the integration of adjacent reachable grids. To sum up, {\GeoRMBR} and {\GeoP} perform much better than {\SpaReach} in initialization even in very dense graphs. {\GeoF} can keep almost the same performance with {\SpaReach} approach.

Figure~\ref{fig:ini_time_go_uniprot} shows the initialization time for all six approaches on the {\gouniprot} dataset. Both {\RMBR} approaches, i.e., {\GeoRMBR} and {\GeoP}, still outperform {\SpaReach}. This is due to the fact that a spatial index constitutes a high proportion of {\SpaReach} initialization time. As opposed to the {\uniprotc} case, the smaller performance gap between initializing {\GeoRMBR} and {\SpaReach} in {\gouniprot}.is explained as follows. The size of \gouniprot is far less than {\uniprotc} which decreases the spatial index initialization cost. As a result, the index construction time in {\SpaReach} is less than that in {\uniprotc}. Since this graph has more reachability information, all {\GeoReach} approaches require more time than in {\uniprotc}. It is conjunction of {\GeoReach} and {\SpaReach} index size changes that causes the smaller gap.

\section{Conclusion}
\label{sec:conclusion}

This paper describes {\GeoReach} a novel approach that evaluates graph reachability queries and spatial range predicates side-by-side. {\GeoReach} extends the functionality of a given graph database management system with light-weight spatial indexing entries to efficiently prune the graph traversal based on spatial constraints. {\GeoReach} allows users to tune the system performance to achieve both efficiency and scalability. Based on extensive experiments, we show that {\GeoReach} can be scalable and query-efficient than existing spatial and reachability indexing approaches in relatively sparse graphs. Even in rather dense graphs, our approach can outperform existing approaches in storage overhead and initialization time and still achieves faster query response time. In the future, we plan to study we plan to study the extensibility of {\GeoReach} to support different spatial predicates. Furthermore, we aim to extend the framework to support a distributed system environment. Last but not least, we also plan to study the applicability of {\GeoReach} to various application domains including: Spatial Influence Maximization, Location and Social-Aware Recommendation, and Location-Aware Citation Network Analysis.

\begin{small}
\bibliographystyle{abbrv}
\bibliography{GraphDB,SpatialIndexing,ReachabilityIndexing,reference,BookIndustryWeb}
\end{small}

\end{document}